\documentstyle[prd,preprint,aps]{revtex}

\begin{document}

% *** Title Page **************************************************************
\title{
		     Model Independent Determination of the         \\
		     Solar Neutrino Spectrum with and without MSW
}
\author{             Naoya Hata\cite{Present-address} and Paul Langacker   }
\address{
{\it
                     Department of Physics,
                     University of Pennsylvania,               \\
                     Philadelphia, Pennsylvania 19104          \\
}}

                     \date{September 20, 1994, UPR-0625T, hep-ph/9409372}

\maketitle
%
% *** Abstract ****************************************************************

\renewcommand{\baselinestretch}{1.3}
\begin{abstract}

    Besides the opportunity for discovering new neutrino physics, solar
    neutrino measurements provide a sensitive probe of the solar interior,
    and thus a rigorous test of solar model predictions.  We present model
    independent determinations of the neutrino spectrum by using relevant
    flux components as free parameters subject only to the luminosity
    constraint.  (1) Without the Mikheyev-Smirnov-Wolfenstein (MSW)
    effect, the best fit for the combined data is poor.  Furthermore, the
    data indicate a severe suppression of the $^7$Be flux relative to the
    $^8$B, contradicting both standard and nonstandard solar models in
    general; the $pp$ flux takes its maximum value allowed by the
    luminosity constraint.  This pathology consistently appears even if we
    ignore any one of the three data.  (2) In the presence of the
    two-flavor MSW effect, the current constraint on the initial $^8$B
    flux is weak, but consistent with the SSM and sufficient to exclude
    nonstandard models with small $^8$B fluxes.  No meaningful constraint
    is obtained for the other fluxes.  In the future, even allowing MSW,
    the $^8$B and $^7$Be fluxes can be determined at the $\pm$(15 -- 20)\%
    level, making competing solar models distinguishable.  We emphasize
    that the neutral current sensitivity for $^7$Be neutrinos in BOREXINO,
    HELLAZ, and HERON is essential for determining the initial fluxes.
    The constraints on the MSW parameters in the model independent
    analysis are also discussed.

\end{abstract}
\pacs{PACS numbers: 96.60.Kx, 12.15.Fp, 14.60.Gh}
%
% *** Text ********************************************************************

\section{		Introduction	}
\label{sec:intro}

The solar neutrino deficit, confirmed by all existing experiments,
challenges our understanding of the Sun as well as of neutrinos.  The
purpose of this paper is to consider the possibility of model
independent determinations of the principle neutrino flux components
from the solar neutrino data, both at present and in the future and
with and without new neutrino properties.  These fluxes can then be
compared with the prediction of any solar model, standard or
nonstandard.  In fact, if the present experiments
\cite{Homestake,Homestake-update,Kamiokande-II,Kamiokande-III,SAGE,%
SAGE-update,GALLEX} are correct, purely astrophysical explanations for
the flux deficit are highly unlikely:
\begin{itemize}

\item	The standard solar models (SSMs) \cite{Bahcall-Pinsonneault,%
	Turck-Chieze-Lopes} are excluded by the data as summarized in
	Table~\ref{tab:expdata}.  The discrepancy cannot be reconciled
	by simply changing input parameters in 	the SSM calculations
	\cite{Bahcall-Bethe}.

\item	The lower observed rate of Homestake relative to Kamiokande
	is incompatible with astrophysical solutions
	\footnote{
		By astrophysical solutions, we include those involving
		nuclear reaction cross sections in the Sun, but not
		the chlorine and gallium detector cross sections.  }
	in general.  This is a much more serious difficulty than the
	simple deficit of observed neutrinos relative to the SSM
	expectations.
	\footnote{%
	    	In fact, the discrepancy in the relative rate is
		{\it aggravated} in models in which the $^8$B flux is
		reduced (e.g., by lowering the core temperature or
		reducing the $^7{\rm Be}(p, \gamma)^8{\rm B}$ cross
		section) to explain the Kamiokande data.}
	A model independent analysis \cite{HBL} suggests a complete
	elimination of the $^7$Be flux and, in addition, a larger
	depletion of the $^8$B spectrum at lower energies and/or
	additional neutral current events from $\nu_\mu$ or $\nu_\tau$
	in Kamiokande.  The larger suppression of the $^7$Be than the
	$^8$B flux contradicts nonstandard solar models in general,
	including ad hoc ones.  A distortion of the $^8$B energy
	spectrum cannot be caused by astrophysical effects at the
	observable level \cite{Bahcall-spectrum}.  $\nu_\mu$ and
	$\nu_\tau$ can interact through the neutral currents in
	electron scattering in Kamiokande, and their existence in the
	solar flux signifies neutrino flavor oscillations.

\item	The problem of the larger suppression of the $^7$Be flux relative
	to $^8$B remains even if we ignore any one of the three data.
	In this sense, the data are consistent with each other.  In
	particular, if we consider the Kamiokande and the gallium
	results only, the nonstandard solar models consistent with the
	Kamiokande result generally predict a gallium rate larger than
	100 SNU,
	\footnote{%
		SNU (solar neutrino unit) = 1/ 10$^{36}$ atoms /sec.}
	inconsistent with the combined result of SAGE and GALLEX
	($77 \pm 9$ SNU).

\end{itemize}

With standard neutrino physics, the current situation forces us to
consider a serious problem with two or more of the experiments {\em
and} a drastic revision of the SSM calculation unless all of the
experiments are wrong.  The Mikheyev-Smirnov-Wolfenstein (MSW)
mechanism \cite{MSW}, on the other hand, provides a complete
description of the data and is also consistent with the SSM (see
\cite{HL-MSW-analysis} and references therein).  Because of the
consistency with the experiments and the simplicity of the theory, we
consider the two-flavor MSW solutions as the most attractive scenario
among many proposed particle physics solutions.

For solar astronomy, whether new neutrino physics is present or not,
the central issue is the determination of the solar neutrino spectrum.
The theory of the Sun, which is the best measured main sequence star,
is the keystone of our understanding of stellar structure and
evolution.  Solar neutrinos are a direct, sensitive probe of the solar
core, and the neutrino flux measurements provide an opportunity for
rigorous tests of solar models, standard or nonstandard.

For the SSM, the neutrino spectrum is a diagnostic of the underlying
assumptions in the theory.  The flux prediction depends on the input
physics, such as the opacity calculation and the nuclear cross
sections, whose uncertainties might be underestimated.  In particular,
the $p(^7{\rm Be},{}^8{\rm B})\gamma$ cross section, which is directly
proportional to the $^8$B flux, was recently measured using the
Coulomb dissociation method in the RIKEN experiment \cite{RIKEN}.
Although the measurement uncertainty is still large, the preliminary
result suggests the cross section can be 25\% lower than the current
standard value \cite{Johnson-etal}.  The SSM also includes
simplifications such as the omission of rotations, magnetic fields,
and the gravitational settling of various elements.  Those effects on
the neutrino flux have never been quantified in the SSM uncertainties.

The nonstandard solar models, most of which are constructed to explain
the solar neutrino deficit, assume nonstandard input parameters or
nonstandard mechanisms.  Examples are the low central temperature
($T_C$), low opacity, low Z, large $S_{11}$,
\footnote{
	$S_{11}$, $S_{33}$, $S_{34}$, and $S_{17}$ are the S factors
	proportional to the cross sections for
	$p+p \rightarrow {}^2{\rm H} + e^+ + \nu_e$,
	${}^3{\rm He} + {}^3{\rm He} \rightarrow {}^4{\rm He} + 2p$,
	${}^3{\rm He} + {}^4{\rm He} \rightarrow {}^7{\rm Be} + \gamma$, and
	$p + {}^7{\rm Be} \rightarrow {}^8{\rm B} + \gamma$, respectively. }
large $S_{33}$, small $S_{34}$, small $S_{17}$, mixing, and weakly
interacting massive particle (WIMP) models.  The neutrino data should
test the validity of such (often ad hoc) assumptions.

To determine the solar neutrino spectrum from the experiments, one
needs to extract from the data the magnitude of the flux, component by
component.  The Kamiokande experiment measures the $^8$B flux
exclusively.  The radio-chemical detectors measure the flux components
only as a weighted sum according to the energy dependence of the
detector cross sections: the Homestake chlorine experiment is
sensitive mainly to the $^8$B flux, but also to the $^7$Be, CNO, and
$pep$ fluxes; the gallium experiments measure all components,
including the dominant $pp$ flux.  In the future, the Sudbury Neutrino
Observatory (SNO) \cite{SNO}, Super-Kamiokande \cite{Super-Kamiokande}
and ICARUS \cite{ICARUS} will measure the $^8$B flux with a high
precision.  BOREXINO \cite{BOREXINO} will be capable of measuring the
$^7$Be line spectrum.  HELLAZ \cite{HELLAZ}, and HERON \cite{HERON}
will observe the $^7$Be flux and the main $pp$ flux individually.

This solar neutrino spectroscopy can be complicated if new particle
physics effects are present, since those effects are often energy
dependent and therefore distort the energy spectrum.  Uncertainties in
the neutrino parameters contribute to uncertainties in the neutrino
flux, and vice versa.  In the presence of the MSW effect, for example,
the determination of the initial (undistorted) flux components
requires a knowledge of the neutrino parameters and, in turn, the
determination of the neutrino parameters depends on the initial flux
magnitudes.

To extricate the neutrino flux components from the data and
distinguish various competing solar models, it is best to consider a
simple and general theoretical framework including all standard and
nonstandard solar models.  Such an analysis scheme should be viable
with and without particle physics effects.

Variations of solar models have usually been considered in model
dependent frameworks.  Monte Carlo SSMs
\cite{Bahcall-Ulrich,Bahcall-book,Bahcall-Bethe} were obtained from various
input parameters normally distributed about their most probable
values.  Those solar models are, however, calculated within the SSM
and do not address the possibility of nonstandard processes omitted in
the modeling or the possibility of input parameters grossly different
from the standard values.  The low $T_C$ model
\cite{BKL,BHKL,Castellani-etal-PRD} parameterizes the neutrino fluxes by
nonstandard core temperatures as power laws
\cite{Bahcall-Ulrich,Bahcall-book}.  The description is more general
than the Monte Carlo SSMs since it includes a large class of
nonstandard solar models.  Again, however, the $T_C$ description is
model dependent: there are nonstandard solar models that cannot be
parametrized simply by nonstandard $T_C$, such as those with
nonstandard $S_{17}$ or $S_{34}$ values.

In this paper, we consider a model independent description of solar
models, characterized by the magnitude of each of the neutrino flux
components.  By setting up an analysis scheme as general as possible,
we depart from particular theoretical constraints.  We hope that the
experiments will distinguish standard and nonstandard models and
eventually identify the correct solar model.  The purpose of this
paper is to demonstrate that such a description is feasible and is a
powerful tool in analyzing the solar neutrino data, especially once
the high precision data from the next generation experiments are
available.

Our model independent analysis originates in Ref.~\cite{HBL} (see also
Ref.~\cite{Castellani-Degl'Innocenti-Fiorentini-AA,Castellani-etal-PRD}).
Here we elaborate the analysis and extend it to the case in which the
two-flavor MSW effect is present.  (Of course, the analysis can be
generalized in the presence of any particle physics effect.)  We
consider magnitudes of the four prominent flux components, $pp$,
$^7$Be, $^8$B, and CNO (the sum of $^{13}$N and $^{15}$O), as free
parameters in fitting data.  In doing so, we make minimal assumptions:

\begin{itemize}

\item   The Sun is in a quasi-static state, and the solar luminosity
        is generated by the ordinary nuclear reactions of
        the $pp$ and CNO chains.  This imposes a relation among the
        fluxes:
	\begin{equation}
	\label{eq:luminosity}
          \phi(pp) + \phi(pep) + 0.958 \, \phi(\mbox{Be})
          + 0.955 \,  \phi(\mbox{CNO})
          = 6.57 \times 10^{10} \, \mbox{cm}^{-2}\mbox{s}^{-1},
	\end{equation}
	where $\phi(\mbox{CNO})$ is the sum of the $^{13}$N and $^{15}$O
	fluxes, which are varied with the same scale factor.

\item   Astrophysical mechanisms do not distort the shape of the energy
        spectrum of the individual flux component at the observable level.
	It was shown that possible distortions of the spectrum due to
	such astrophysical effects as gravitational red-shifts and
	thermal fluctuations are completely negligible
	\cite{Bahcall-spectrum}.  On the other hand, particle physics
	effects such as the MSW mechanism are in general energy dependent
	and lead to significant spectral distortions.

\item   The detector cross section calculations \cite{Bahcall-Ulrich,%
	Bahcall-book,Aufderheide-etal} are correct.

\item	The minor fluxes ($pep$
		\footnote{
		The $pep$ neutrinos are from the reaction
		$p+e^-+p \rightarrow {}^2\mbox{H} + \nu_e$.
		The $pep$ flux is the largest among the three minor
		fluxes.  It is strongly correlated with the $pp$ flux
		in many of the nonstandard solar models
		\cite{Bahcall-Ulrich-nonssm-table} and does not vary
		significantly from model to model.  Of course, one can
		also use the $pep$ flux as a free parameter.},
	$^{17}$F, and $hep$
		\footnote{%
		The $hep$ neutrinos are from the reaction
		${}^3\mbox{He} + p \rightarrow {}^4\mbox{He} + e^+ + \nu_e$.}
	) are set to the SSM values.

\end{itemize}

The simultaneous (global) analysis of all data is essential in
obtaining constraints on fluxes.  No one experiment provides enough
information to determine the entire neutrino spectrum; only by
combining various experiments with different energy thresholds is it
possible to determine each flux component and test solar model
predictions.  For example, if there are no new particle physics
effects, then by combining the $^8$B flux measured in Kamiokande with
the Homestake and gallium experiments, one can deduce an absence of
the $^7$Be flux and a detection of the $pp$ flux.  Allowing for MSW or
other particle physics effects, a global analysis is even more
essential because one must simultaneously determine the initial fluxes
and the MSW-induced flux reduction and spectral distortions.

In constraining the fluxes and testing solar models, a consistent
joint analysis is important.  Taking overlaps of parameter space
allowed by different experiments does not yield a correct estimation
of uncertainties \cite{HL-MSW-analysis}; one needs to carry out proper
joint $\chi^2$ analyses, which are essentially identical to the
maximum likelihood method in gaussian cases.  Since we are testing
theoretical models statistically, the experimental and theoretical
uncertainties have to be incorporated; a proper treatment of the
correlations among uncertainties is also important
\cite{HL-MSW-analysis}.

When the MSW effect is present, it is best to incorporate available
energy spectrum and day-night asymmetry data to obtain additional
constraints on the neutrino parameters and therefore on the fluxes.
Those constraints from future high-counting experiments would be
especially useful.  In this paper, however, we do not incorporate
the existing energy spectrum and day-night data from Kamiokande
\cite{Kamiokande-II}; their uncertainties are large and do not
significantly change the results obtained from the averaged data.
\footnote{
	Also, Kamiokande has only presented the spectrum and angular
	(direction of the Sun with respect to the nadir) distributions
	separately.  Since they are based on the same data, the two
	distributions are correlated and cannot be used
	simultaneously.  It is recommended that in the future the data
	be presented in bins of definite energy {\it and} angle.}

The rest of the paper is organized as the following.  In
Section~\ref{sec:without-MSW}, the constraints on the fluxes without
introducing particle physics effects are obtained.  The joint fit of
the combined the Homestake, Kamiokande, SAGE, and GALLEX data yields a
poor $\chi^2$ value.  The best fit is, in fact, obtained for a
negative $^7$Be flux, suggesting a serious problem with the
experiments {\em or} the existence of new particle physics effects: a
distortion of the $^8$B spectrum and/or neutral current contributions
from $\nu_\mu$ or $\nu_\tau$ in Kamiokande.  Even if we accept this
poor fit assuming standard neutrinos, the constraint on the fluxes
contradicts nonstandard solar models in general.  The constraints on
the fluxes will be displayed in the $^7$Be--$^8$B, $pp$--$^7$Be, and
$pp$--$^8$B planes.  (For simplicity, we refer to neutrino fluxes in
units of the reference fluxes listed in Table~\ref{tab:BP-fluxes}
unless otherwise mentioned.  Those fluxes correspond to the
Bahcall-Pinsonneault fluxes with the helium diffusion effect
\cite{Bahcall-Pinsonneault}).  The results when one of the three
experiments are omitted will be also given.  By ignoring one
experiment, the uncertainties in the flux constraints become larger,
but the constraints are consistent with those obtained from all data,
and again contradict astrophysical/nuclear solutions.

In Section~\ref{sec:without-MSW}, we also discuss possible constraints
on the fluxes from the next generation experiments.  In fact, if no
new particle effects are present, SNO and Super-Kamiokande will
determine the initial $^8$B flux with a high precision, and BOREXINO,
HELLAZ, and HERON will measure the initial $^7$Be flux exclusively.
The flux constraints from hypothetical results from these experiments
with various central values and various measurement uncertainties will
be examined.

In Section~\ref{sec:with-MSW}, we consider the constraints on the
fluxes when the two-flavor MSW effect for transitions $\nu_e \rightarrow
\nu_\mu$ or $\nu_e \rightarrow \nu_\tau$ is present.  Our analysis scheme
can in principle be applied to other particle physics scenarios, such
as three-flavor MSW, transitions into sterile neutrinos, vacuum
oscillations, a large neutrino magnetic moment, neutrino decays,
flavor changing neutral currents, violation of the equivalence
principle, etc.  We consider the two-flavor MSW solution because of
its simplicity and viability.  It is likely that if two-flavor MSW is
indeed occurring, there will be enough complementary information
[e.g., from spectral distortions, day-night asymmetries, and SNO
neutral current (NC) measurements] to establish it as the most likely
candidate even allowing nonstandard solar models
\cite{HL-MSW-analysis}.  Of course one could never rigorously exclude
the possibility of more complicated scenarios, such as the
simultaneous importance of transitions into $\nu_\mu$ (or $\nu_\tau$)
and sterile neutrinos, which would interfere with the model
independent flux determinations.  For this, one must invoke Occam's
razor.

Once the MSW parameters are introduced as additional free parameters
in the joint fit, constraining the fluxes from the data is not
trivial.  The MSW effect can distort the energy spectrum depending on
the parameters, and can change the contribution from different flux
components.  With the existing data, we can constrain the $^8$B flux
only roughly. Even though the chlorine and gallium experiments have a
sensitivity to the $^7$Be flux, the survival probability of the flux
can be zero for the MSW small-angle (nonadiabatic) solution, and no
meaningful constraint is obtained for the $^7$Be flux.  To constrain
the fluxes and the MSW parameters simultaneously, we need results from
the future experiments, especially the neutral current measurement in
SNO and the $^7$Be neutrino measurement in BOREXINO, HELLAZ, or HERON.
The neutrino-electron scattering mode in these $^7$Be measurements has
a sensitivity to the neutral current interactions with $\nu_\mu$ and
$\nu_\tau$, whose cross sections are 21\% of $\nu_e$'s at this energy.
We will present the possible constraints assuming various outcomes
from those experiments, and show that such a model independent
analysis can determine the solar neutrino spectrum with an accuracy
sufficient to test solar model predictions.  We note that our choice
of the hypothetical results from the SNO NC and BOREXINO experiments
are minimal; additional information from the SNO charged current (CC)
rate, the Super-Kamiokande rate, and the spectral and day-night
asymmetry measurements in SNO and Super-Kamiokande should make the
constraints even better.

\section{	Flux Constraints Assuming Standard Neutrinos	}
\label{sec:without-MSW}

\subsection{ 		Present					}

We consider the constraints on the neutrino fluxes from the updated
solar neutrino data listed in Table~\ref{tab:expdata}.  The main
results are displayed in the $^7$Be--$^8$B plane (although some of the
results are also shown in the $pp$--$^7$Be and $pp$--$^8$B planes).
When the data are fit, the the pp and CNO fluxes are varied freely for
each $\phi({\rm Be})$ and $\phi({\rm B})$, subject only to the
luminosity constraint.  This representation in the $^7$Be--$^8$B plane
is effective since it can display every possible solar model, standard
or nonstandard, that satisfies our minimal assumptions.  Since
predictions for those fluxes vary substantially from model to model,
the $^7$Be--$^8$B plane also provides a useful diagnostic for
experimentally distinguishing competing solar models.

\begin{figure}[h]
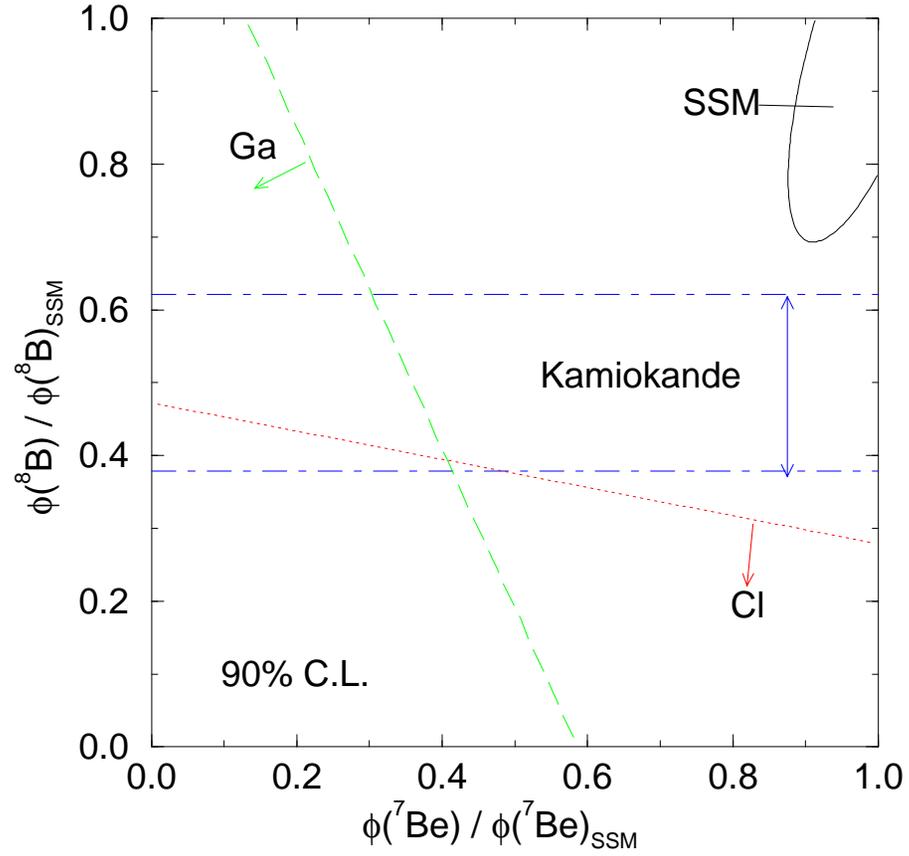


\caption{
The constraints on the $^7$Be and $^8$B fluxes when the Kamiokande,
Homestake, and the combined gallium results are fit separately.  For
each point in this plane, the data are fit to the $pp$ and CNO fluxes
subject to the luminosity constraint.  The fluxes allowed by the
Homestake and gallium result are below the dotted and dashed line,
respectively.  The fluxes allowed by Kamiokande is between the
dot-dashed lines. }
\label{fig:fff_each}
\end{figure}

When the Kamiokande, Homestake, and the combined gallium experiments
of SAGE and GALLEX are fit separately, the constraints on the $^7$Be
and $^8$B fluxes are shown in Fig.~\ref{fig:fff_each}.  The fits
include the uncertainties in the radio-chemical detector cross
sections and in the minor fluxes, which are set to the SSM values.
The Kamiokande result constrains the $^8$B flux; the Homestake result
constrains the $^7$Be, $^8$B, and CNO fluxes; the gallium results
constrain all fluxes including the $pp$.

\begin{figure}[h]
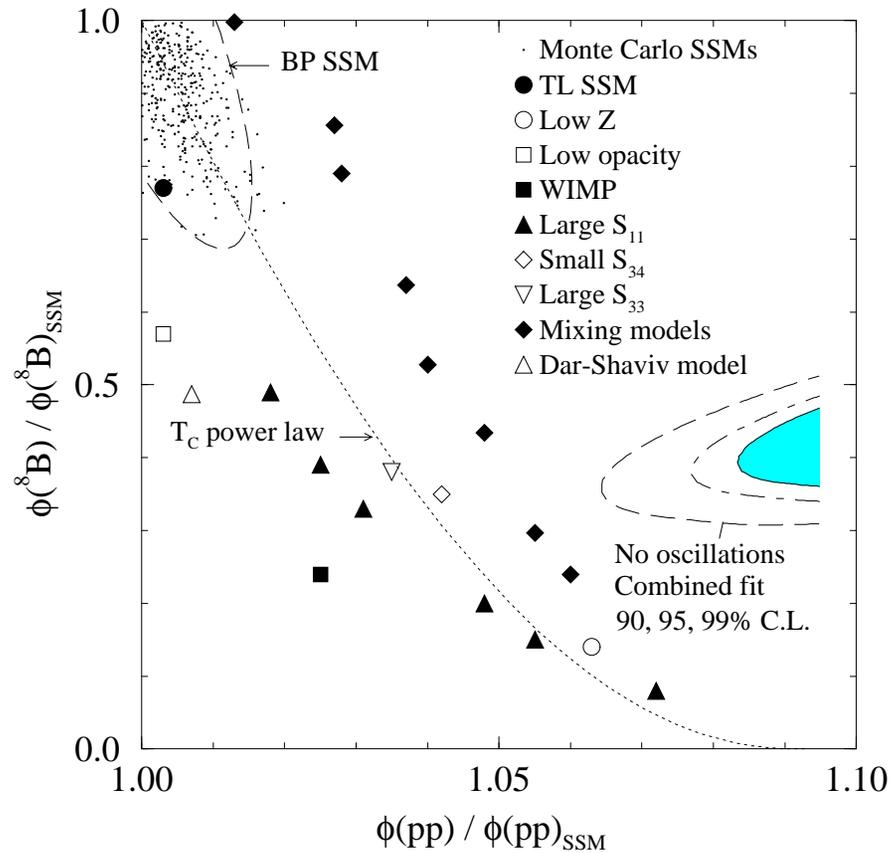


\caption{
The flux constraints obtained from the combined Kamiokande, Homestake,
and gallium results.  The constraints are shown for the (a)
$^7$Be--$^8$B, (b) $pp$--$^7$Be, and (c) $pp$--$^8$B planes.  The best
fit parameters are $\phi(pp)/\phi(pp)_{\rm SSM} = 1.095$, $\phi({\rm
Be})/\phi({\rm Be})_{\rm SSM} = 0$, $\phi({\rm B})/\phi({\rm B})_{\rm
SSM} = 0.41$, and $\phi({\rm CNO})/\phi({\rm CNO})_{\rm SSM} = 0$
(Table~\protect\ref{tab:flux-constraints}), but this fit is poor:
$\chi^2_{\rm min} / 1 \; {\rm d.f.} = 3.3$, which is excluded at 93\%
C.L.  Also displayed are the Bahcall-Pinsonneault SSM 90\% region
(BP-SSM) \protect\cite{Bahcall-Pinsonneault}, the Bahcall-Ulrich Monte
Carlo SSMs \protect\cite{Bahcall-Ulrich}, the Turck-Chi\`eze--Lopes
(TL) SSM \protect\cite{Turck-Chieze-Lopes}, and various nonstandard
solar models (see the text).  The observations are inconsistent with
any of those standard and nonstandard solar models.  Smaller $S_{17}$
values decrease only the $^8$B flux [as indicated by the downward
arrow in (a)], and {\it aggravate} the discrepancy between the
combined data and nonstandard solar models.  }
\label{fig:curr_all}
\end{figure}

When all data are fit simultaneously, the allowed fluxes are severely
constrained, as shown in Fig.~\ref{fig:curr_all}.  The best fit for
physical (i.e., non-negative) fluxes are obtained for zero $^7$Be and
CNO fluxes, and the $^8$B flux is about 40\% of the SSM prediction;
the absence of the $^7$Be and CNO fluxes forces the $pp$ flux to be
the maximum value (1.095 SSM) allowed by the luminosity constraint
(Eq.~\ref{eq:luminosity}).  These constraints at 1$\sigma$
uncertainties are summarized in Table~\ref{tab:flux-constraints}; they
are also listed as absolute fluxes in
Table~\ref{tab:flux-constraints-abs}.  This model independent result
displays serious problems for any purely astrophysical explanation for
the solar neutrino deficit \cite{HBL}:
\begin{itemize}

\item 	The best fit is poor; the
	$\chi^2$ minimum is in fact obtained for the unphysical value
	$\phi({\rm Be})/\phi({\rm Be})_{\rm SSM} = -0.5$.
	Imposing positivity of the flux, $\chi^2_{\rm min} = 3.3 /
	1\; {\rm d.f.}$,
	\footnote{%
		The fit is in fact for 0 d.f.\ [3 data - (4 parameters
		- 1 constraint)].  For $\chi^2$ values other than
		zero, there is no standard statistical interpretation
		exists other than to conclude that this model is
		excluded.  To quantify the confidence level, we
		allow 1 d.f.\ by considering that the $^7$Be flux is
		fixed to zero. 	}
	which is excluded at 93\% confidence level (C.L.)
	That is, any possible solar model explanation consistent with
	our minimal assumptions is excluded at least at the 93\% C.L.

\item	Even if one accepts this poor fit, the allowed
	fluxes are difficult to explain.  Since $^8$B is produced through the
	reaction $p + {}^7{\rm Be} \rightarrow {}^8{\rm B} + \gamma$, any
	reduction in $^7$Be causes at least an equal reduction in
	$^8$B.  Therefore, unless there is some independent mechanism to
	suppress only the $^7$Be flux
	\footnote{
	For example, both $\phi({\rm Be})$ and $\phi({\rm B})$ could be
	suppressed by a low $T_C$, and $\phi({\rm B})$ could then be enhanced
	by a {\it larger} $S_{17}$.  However, for any realistic $S_{17}$,
	this enhancement would be negligible.  }
	or the uncertainty in the $^7$Be
	electron capture rate is grossly underestimated, the
	$^8$B flux is expected to be reduced more than the $^7$Be flux,
	contrary to the data.

\item	Finally, various standard and nonstandard models are
	also displayed in Fig.~\ref{fig:curr_all}: the Bahcall-Pinsonneault
	SSM including the helium diffusion effect \cite{Bahcall-Pinsonneault},
	the Bahcall-Ulrich 1000 Monte Carlo SSMs \cite{Bahcall-Ulrich},
	Turck-Chi\`eze--Lopes SSM \cite{Turck-Chieze-Lopes}, the low Z
	model \cite{Bahcall-Ulrich,Bahcall-book}, the low opacity
	models with the opacity reduced by 10 and 20\%  \cite{Dearborn},
	the WIMP model \cite{WIMPs}, the large $S_{11}$ models
	\cite{Castellani-Degl'Innocenti-Fiorentini-largeS11}, the small
	$S_{34}$ model \cite{Turck-Chieze-Lopes}, the large $S_{33}$ model
	\cite{Turck-Chieze-Lopes}, the mixing models \cite{Mixing-model},
	the Dar-Shaviv SSM \cite{Dar-Shaviv}, and the high Y model
	\cite{Bahcall-Ulrich,Bahcall-book}.
	Also shown are models parametrized by a lower $T_C$ (which
	approximately incorporates many of the explicit models) and a
	lower $S_{17}$.  As seen in Fig.~\ref{fig:curr_all}, none of
	those solar model predictions are even close to the observations.

\end{itemize}

We also note that a lower $S_{17}$ value, suggested by the RIKEN
experiment \cite{RIKEN}, aggravates the problem with
astrophysical/nuclear solutions, contrary to the general notion.  A
lower $S_{17}$ value can make the theory prediction for the $^8$B flux
smaller and closer to the Kamiokande result, which leaves little room
to introduce other astrophysical/nuclear effects (e.g., a lower $T_C$)
to reduce the $^7$Be flux, failing to explain either the Homestake or
the gallium results.

This complete phenomenological failure of astrophysical solutions
suggests nonstandard particle physics effect such as the MSW effect,
or serious problems with the experiments \cite{HBL}.

\begin{figure}[h]
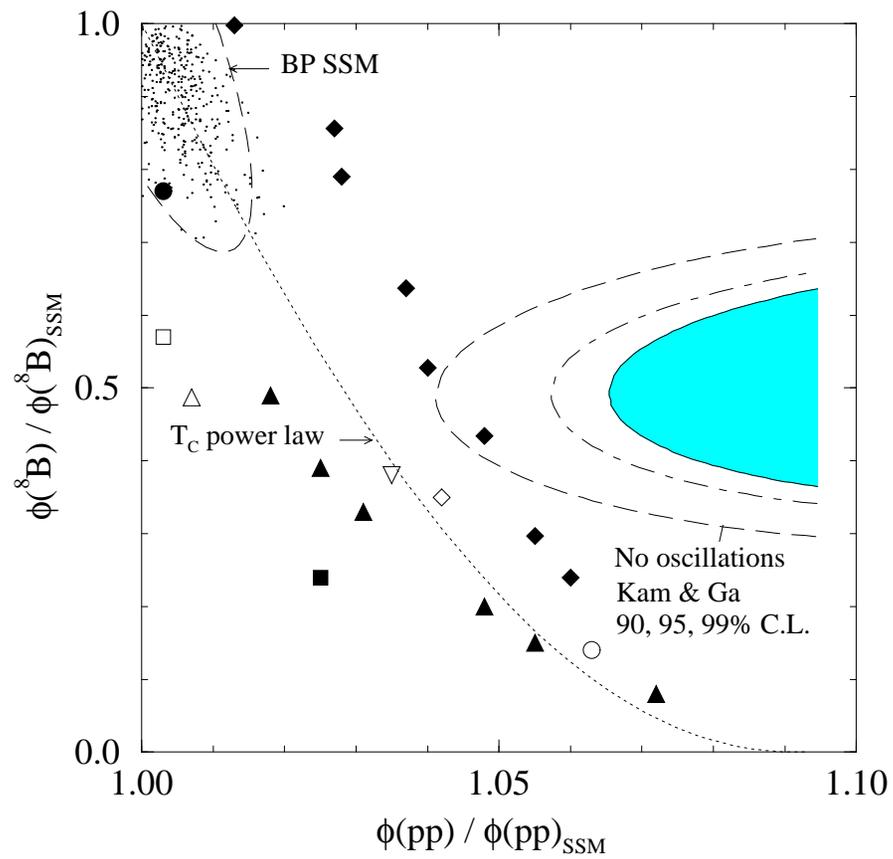


\caption{
The flux constraints from the Kamiokande and gallium data, but without
the Homestake result.  The combined fit again indicates the larger
suppression of the $^7$Be flux relative to $^8$B, consistent with the
constraint including the Homestake result
(Fig.~\protect\ref{fig:curr_all}).  The C.L.\ contours in (a)
correspond to $\phi({\rm CNO}) = 0$, while the nonstandard models
within the 99\% C.L. (the mixing models, the large $S_{33}$ model, and
the small $S_{34}$ model) predict non-zero CNO fluxes, aggravating the
disagreement with the data.}
\label{fig:curr_no-cl}
\end{figure}

\begin{figure}[h]
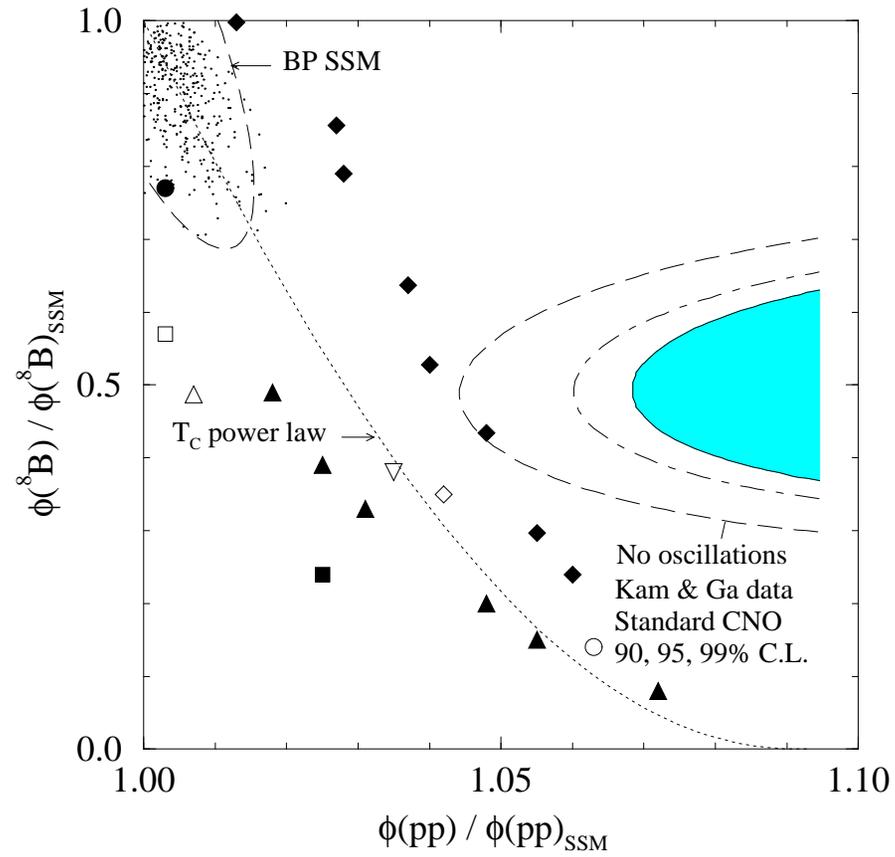


\caption{
The flux constraints from the Kamiokande and gallium results, but
without the Homestake data, when the standard CNO flux is assumed.  A
non-zero CNO flux aggravates the disagreement between the data and
solar model predictions.}
\label{fig:curr_no-cl_cno1}
\end{figure}

Even if only the Kamiokande and gallium results are considered, there
is still essentially no viable theoretical explanation.  Although the
best fit somewhat improves ($\chi^2 / 0\; {\rm d.f.} = 1.2$), the
obtained fluxes displayed in Fig.~\ref{fig:curr_no-cl} are consistent
with a complete depletion of the $^7$Be flux, while the $^8$B flux is
about half of the SSM prediction (Table~\ref{tab:flux-constraints}
and~\ref{tab:flux-constraints-abs}).  This is again in severe
contradiction with nonstandard solar models in general.  The
nonstandard solar models that are significantly inside the 99\% C.L.\
contour in Fig.~\ref{fig:curr_no-cl}(a) are the small $S_{34}$ model,
the large $S_{33}$ model, and ad hoc mixing models that involve a core
with 0.4 and 0.8 solar masses that is mixed continuously.  These
models also predict non-zero CNO fluxes, while the C.L.\ contours in
Fig.~\ref{fig:curr_no-cl} corresponds to zero CNO flux.  The non-zero
CNO contribution further aggravates the disagreement.  When the CNO
flux is fixed to the SSM value, the constraint for the combined
Kamiokande and gallium results is displayed in
Fig.~\ref{fig:curr_no-cl_cno1}.

\begin{figure}[h]
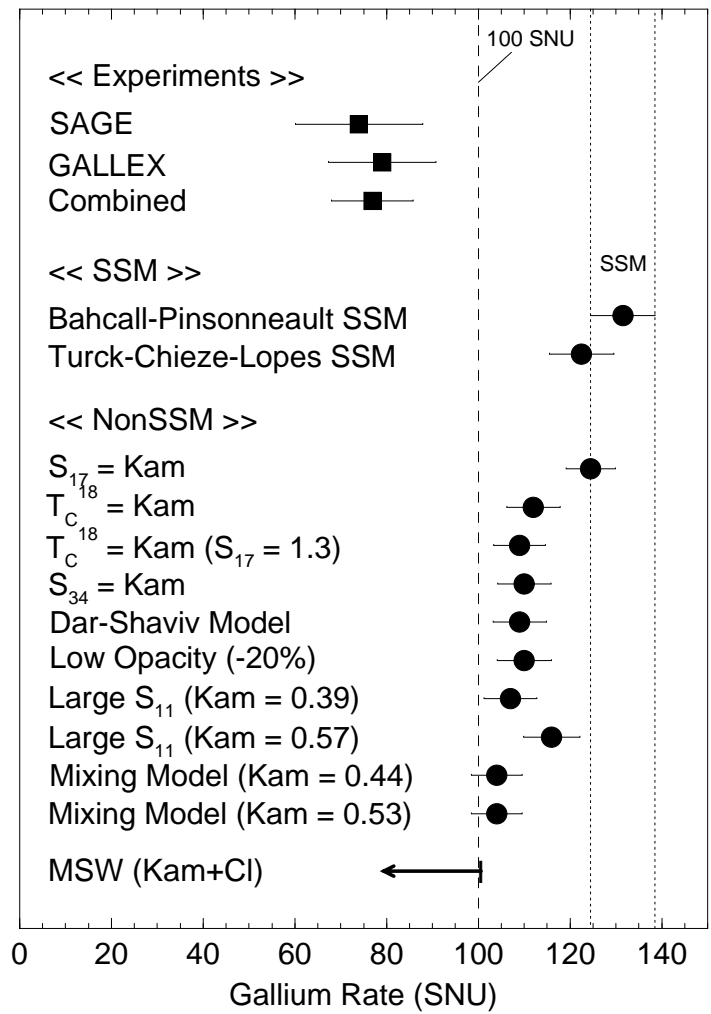


\caption{
The gallium experiment results, the SSM gallium rates, and the gallium
rates of nonstandard solar models which predict the $^8$B flux
consistent with or close to the $^8$B flux observed in Kamiokande (see
the text for details).  The nonstandard solar models consistent with
Kamiokande predict the gallium rate $R_{\rm Ga} \protect\gtrsim 100$ SNU,
contradicting the combined observed rate, $77 \pm 9$ SNU.  The MSW
solution obtained from the combined Kamiokande and Homestake data
predicts $R_{\rm Ga} < 100$ SNU \protect\cite{HL-MSW-analysis},
consistent with the data. }
\label{fig:ga_constraint}
\end{figure}

The discrepancy between solar model solutions and the combined
Kamiokande and gallium result can be described in another way.
Nonstandard models yield a wide variety of fluxes, and therefore a
large range for their gallium predictions: from the 78 SNU of the
luminosity limit
	\footnote{
	This corresponds to zero $^7$Be, CNO, $pep$, and $^8$B fluxes
	and the $pp$ flux with the maximum value allowed by the
	luminosity constraint (Eq.~\ref{eq:luminosity}).  }
to 303 SNU of the maximum rate model \cite{Bahcall-Pinsonneault}.
However, the $^8$B flux, which has the largest uncertainty among the
major fluxes, has been constrained by Kamiokande, and this in turn
constrains the gallium predictions of nonstandard models.  Such a
constraint was considered in the SSM framework with the Monte Carlo
method \cite{Bahcall-Bethe}, but here we allow nonstandard models as
well.  Displayed in Fig.~\ref{fig:ga_constraint} along with the
gallium data are the gallium predictions of various SSMs and also of
nonstandard solar models that are consistent with or close to the
$^8$B flux observed in Kamiokande: the model with $S_{17}$ normalized
to the Kamiokande result, the low $T_C$ model with a reduction of 4\%,
the model with a {\it larger} $S_{17}$ (30\%) and a lower $T_C$ (5\%),
the low $S_{34}$ (50\%) model, the Dar-Shaviv SSM \cite{Dar-Shaviv},
the low opacity model \cite{Dearborn}, the large $S_{11}$ models that
predict $\phi({\rm B})/\phi({\rm B})_{\rm SSM} = 0.39$
\cite{Turck-Chieze-Lopes} and $0.57$
\cite{Castellani-Degl'Innocenti-Fiorentini-largeS11}, and mixing
models \cite{Mixing-model}.  The uncertainties include the $^8$B
uncertainty due to the Kamiokande uncertainty (14\%), but the dominant
contribution is from the gallium cross section uncertainty.  From this
list, we obtain
\begin{equation}
	\mbox{Gallium rate consistent with Kamiokande}
	\gtrsim 100\; {\rm SNU},
\end{equation}
while the combined gallium rate of SAGE and GALLEX is $77 \pm 9\;
\mbox{SNU}$.

The lower limit of 100 SNU can be roughly understood as followings.
The nonstandard solar models considered here all predict smaller
reductions of the $^7$Be than the $^8$B flux, and this, combined with
the Kamiokande result, gives the lower limit on the $^7$Be flux to be
about half of the SSM value, contributing at least 18.3 SNU to the
gallium rate.  The $pp$ and $pep$ fluxes do not depend significantly
on solar models; the luminosity constraint and a decrease in the
$^7$Be flux result in an increase in the $pp$ flux by 5\% (3.7 SNU).
Adding these ($pp$, $pep$, $^7$Be, and $^8$B) gives a total of 102.9
SNU, with uncertainties in the treatment of the CNO fluxes, the
gallium detector cross section, and the $^8$B measurement in
Kamiokande.

This discrepancy is extremely important because it is independent of
the Homestake result, but displays exactly the same symptom as in the
Kamiokande-Homestake comparison: the absence of the $^7$Be flux, for
which astrophysics offers no explanation.  Furthermore, experimental
developments in the near future will significantly influence the
situation.  The calibration of the gallium detectors with chromium
sources will help understand the systematic uncertainty and the
detector cross section, reducing the uncertainty.  It is also
important to continue the gallium experiments to the statistics limit
to establish consistency or inconsistency with the 100 SNU benchmark.
Theoretically, those models which predict 100 SNU can be compared with
helioseismology data.  In fact, some of the nonstandard models (the
low $T_C$ model
\cite{Gough-Toomre}, the large $S_{11}$ model \cite{Turck-Chieze-Lopes},
the mixing model \cite{Merryfield}, and the low Y model
\cite{Gough-Toomre}) are in conflict with the sound
speed profile inferred from helioseismology observations and therefore
excluded.  Further detailed testing of those nonstandard solar models
with helioseismology data would be welcome.

\begin{figure}[h]
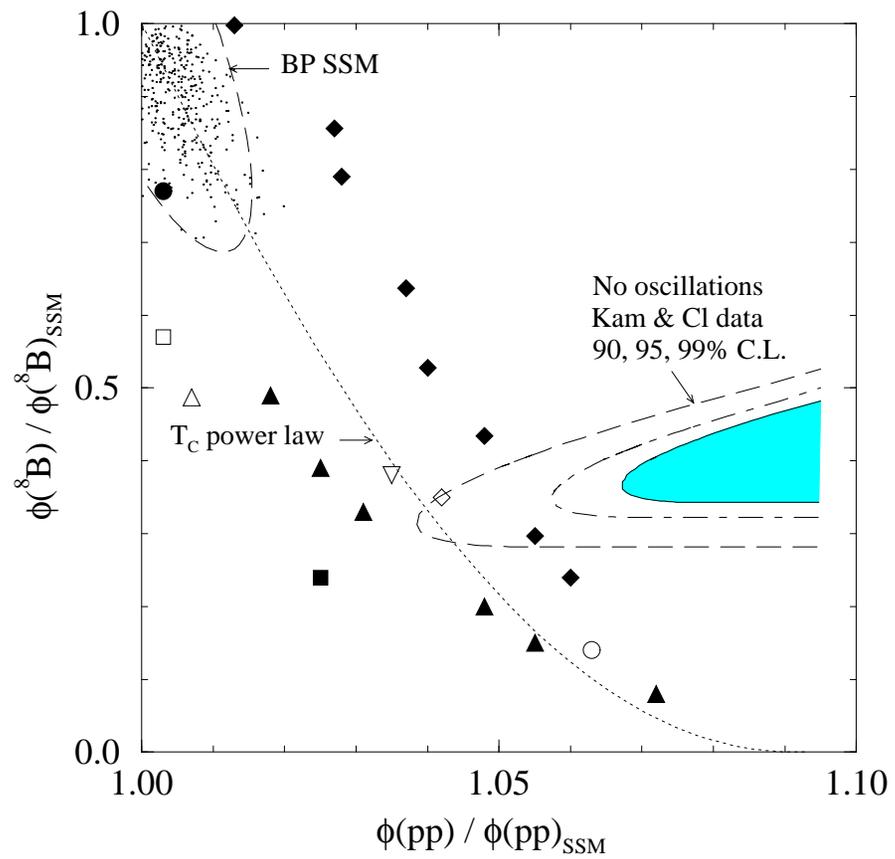


\caption{
The flux constraints from the Kamiokande and Homestake results, but
without the gallium data.  The constraints are consistent with those
including the gallium data (Fig.~\protect\ref{fig:curr_all}).  }
\label{fig:curr_no-ga}
\end{figure}

\begin{figure}[h]
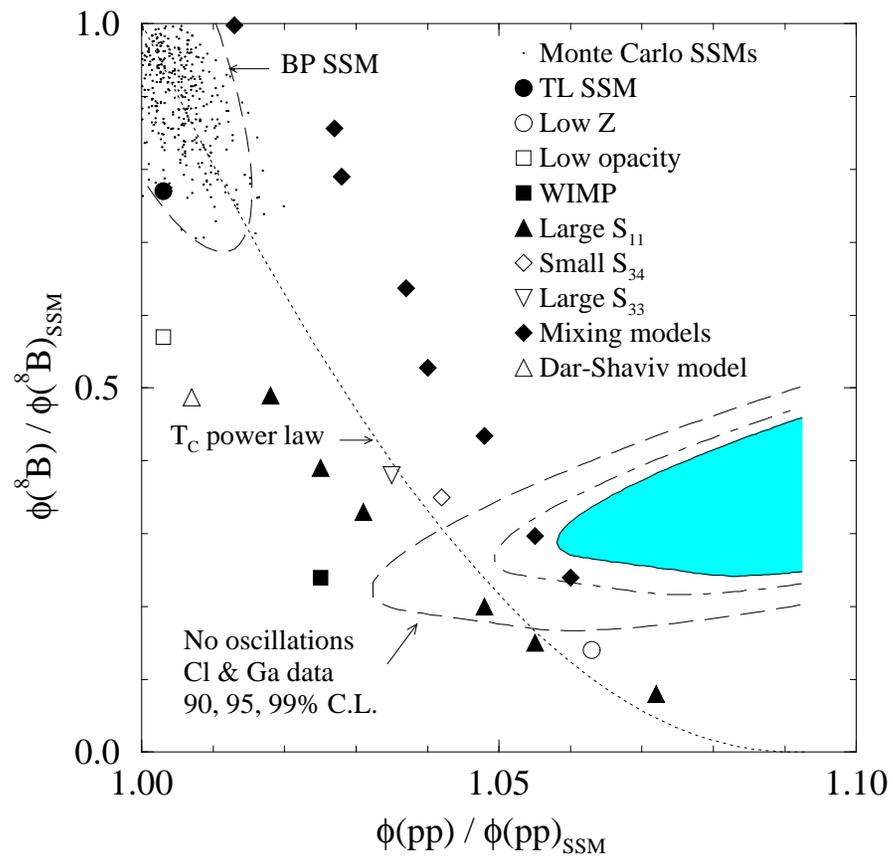


\caption{
The flux constraints from the Homestake and gallium results, but
without the Kamiokande data.  The constraints are consistent with
those including the Kamiokande data (Fig.~\protect\ref{fig:curr_all}).  }
\label{fig:curr_no-kam}
\end{figure}

The flux constraints when the gallium and Kamiokande results are
separately ignored are shown in Figs.~\ref{fig:curr_no-ga} and
\ref{fig:curr_no-kam}.  The flux constraints from various combinations
of the existing data are summarized in
Table~\ref{tab:flux-constraints} and~\ref{tab:flux-constraints-abs}.
We note that any combination of two experiments are consistent with
the complete absence of the $^7$Be and CNO fluxes, the $^8$B flux of
about 40\% of the SSM, and the maximum $pp$ flux, contradicting
astrophysical solutions in general.  That is, we have to ignore two of
the three data to find a reasonable astrophysical explanation of the
solar neutrino problem.

\subsection{ 		Future                                   }

Since the current results are almost limited by systematic
uncertainties, the present status described in the previous section is
unlikely to change with the existing experiments unless there is a
drastic revision in the data analyses.  We expect, on the other hand,
that our understanding of solar neutrinos will greatly improve once
the results from the new generation of high-statistic experiments are
available.  SNO \cite{SNO} and Super-Kamiokande
\cite{Super-Kamiokande} will start in 1996, measuring the $^8$B flux
with high precision.  The neutral current (NC) measurement in SNO and
the measurements of the energy spectrum and time dependence in the two
experiments will either confirm or rule out the neutrino oscillation
hypothesis.  BOREXINO \cite{BOREXINO} will operate later in the decade
and measure the $^7$Be line spectrum separately.  HELLAZ
\cite{HELLAZ} and HERON \cite{HERON} can measure the $pp$ and $^7$Be
neutrinos separately.

\begin{figure}[h]
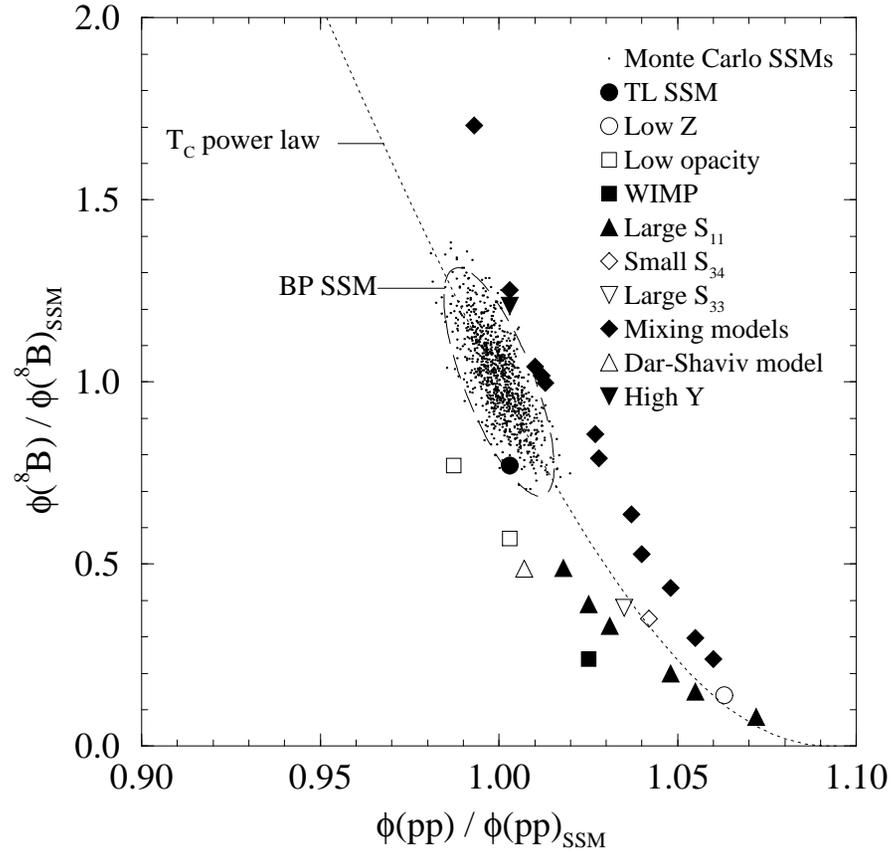


\caption{%
Various standard and nonstandard solar models displayed in the (a)
$^7$Be--$^8$B, (b) $pp$--$^7$Be and (c) $pp$--$^8$B flux parameter
space.  SNO and Super-Kamiokande will measure the $^8$B flux.  The SNO
NC measurement will constrain the $^8$B flux even if neutrino flavor
oscillations are present.  BOREXINO, HELLAZ, and HERON will measure
the $^7$Be flux.  HELLAZ and HERON will also be capable of measuring
the $pp$ flux.  The determinations of the initial $^7$Be and $^8$B
fluxes at the $< 20$\% level will make competing solar models
distinguishable.  For the $pp$ flux, a determination at the few \%
level would be useful.}
\label{fig:nonssms}
\end{figure}

Assuming that neutrino physics effects are absent, we should be able
to calibrate solar models with precision measurements of the $^7$Be
and $^8$B fluxes, independent of the existing experiments.  The
relevant flux parameter space with various standard and nonstandard
solar models is displayed in Fig.~\ref{fig:nonssms}(a).  The
determination of the $^7$Be and $^8$B fluxes at better than the 20\%
level should distinguish between competing solar models.  It is also
important to compare the future results to the present constraints for
a consistency check among the data.

For experiments sensitive to the $pp$ flux, such as the gallium
experiments, HELLAZ, and HERON, the relevant flux parameter space will
be in Fig.~\ref{fig:nonssms}(b) and (c).  The $pp$ flux is, however,
strongly constrained by the solar luminosity, and, to further
distinguish the competing solar models, measurement uncertainties as
small as a few \% will be required.

To study the sensitivity for determining the fluxes and distinguishing
solar models, we have carried out a joint analysis assuming various
possible outcomes from the new generation experiments.  The constraint
on the fluxes is shown in Fig.~\ref{fig:beb_no-msw_each}(a) when SNO
or Super-Kamiokande data are assumed to be $0.50 \pm 0.05$ SSM.  The
allowed parameter space is for 90\% C.L.  The constraint on the $^7$Be
flux from BOREXINO
\footnote{%
	The same analysis applies for the HELLAZ and HERON experiments.}
data ($1.0 \pm 0.1$ SSM) is displayed in
Fig.~\ref{fig:beb_no-msw_each}(b).  Displayed in
Fig.~\ref{fig:beb_no-msw_diff}(a) are the constraints on both fluxes
when the $^7$Be flux is measured in BOREXINO at the SSM value with an
experimental uncertainty of 10\%; various values (0.3, 0.5, 0.7, and
1.0 SSM) for the $^8$B flux measurement are assumed with a 10\%
experimental uncertainty.  Fig.~\ref{fig:beb_no-msw_diff}(b) is the
same except that various $^7$Be values are assumed for a fixed central
value of the $^8$B flux ($0.50 \pm 0.05$ SSM).

The constraints are shown in Fig.~\ref{fig:beb_no-msw_errors} when
different experimental uncertainties are used for the $^8$B flux
measurements and for BOREXINO.  With measurement uncertainties at the
10\% level in SNO, Super-Kamiokande, and BOREXINO, the $^7$Be and
$^8$B fluxes are determined accurately enough that the observations
can distinguish between standard and nonstandard solar models and
perhaps even constrain the SSM parameters.  The constraints from the
future data should be compared with the current constraint
(Fig.~\ref{fig:curr_all}) for a consistency check.

\section{	Flux Constraints Assuming MSW		}
\label{sec:with-MSW}

\subsection{ 		Present                                  }

\begin{figure}[h]
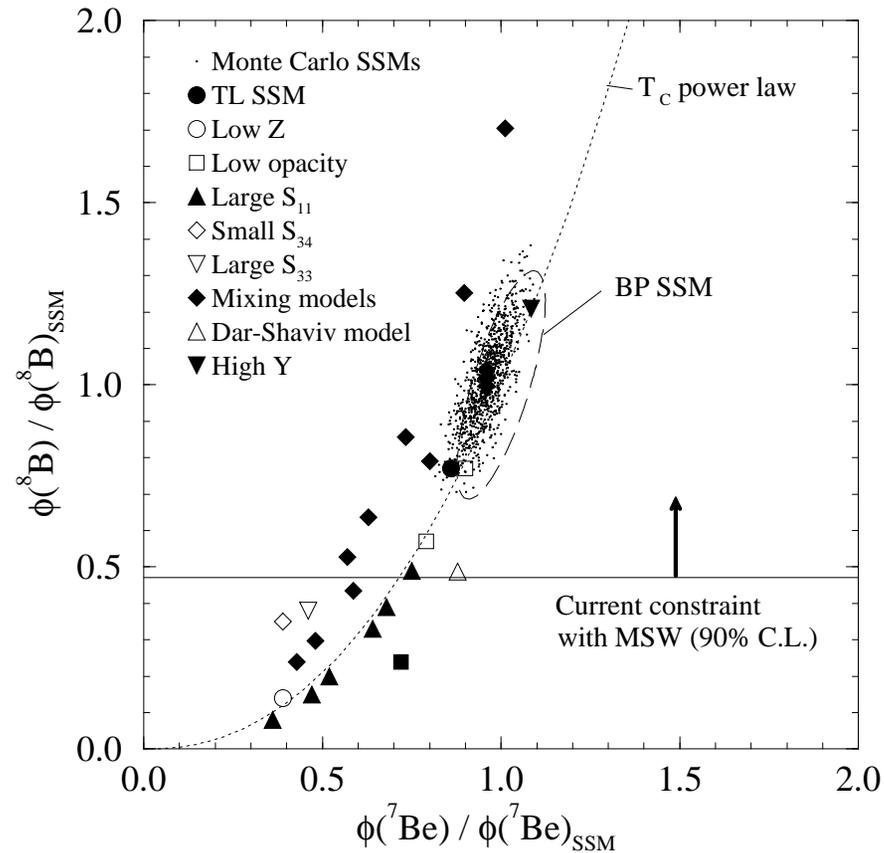


\caption{%
The flux constraint from the existing data when the MSW effect is
assumed.  The current data constrain $\phi({\rm B})/\phi({\rm B})_{\rm
SSM} = 0.47 - 2.07$ (90\% C.L.) as shown in
Fig.~\protect\ref{fig:Bfree_msw} (a).  The solar models with too small
$^8$B fluxes are inconsistent with the existing data and the MSW
hypothesis.  The corresponding allowed MSW parameter space is
displayed in Fig.~\protect\ref{fig:Bfree_msw} (b).  No reasonable
constraint is obtained when the fluxes other than $^8$B are used as
free parameters.  }
\label{fig:beb_curr_msw}
\end{figure}

\begin{figure}[h]
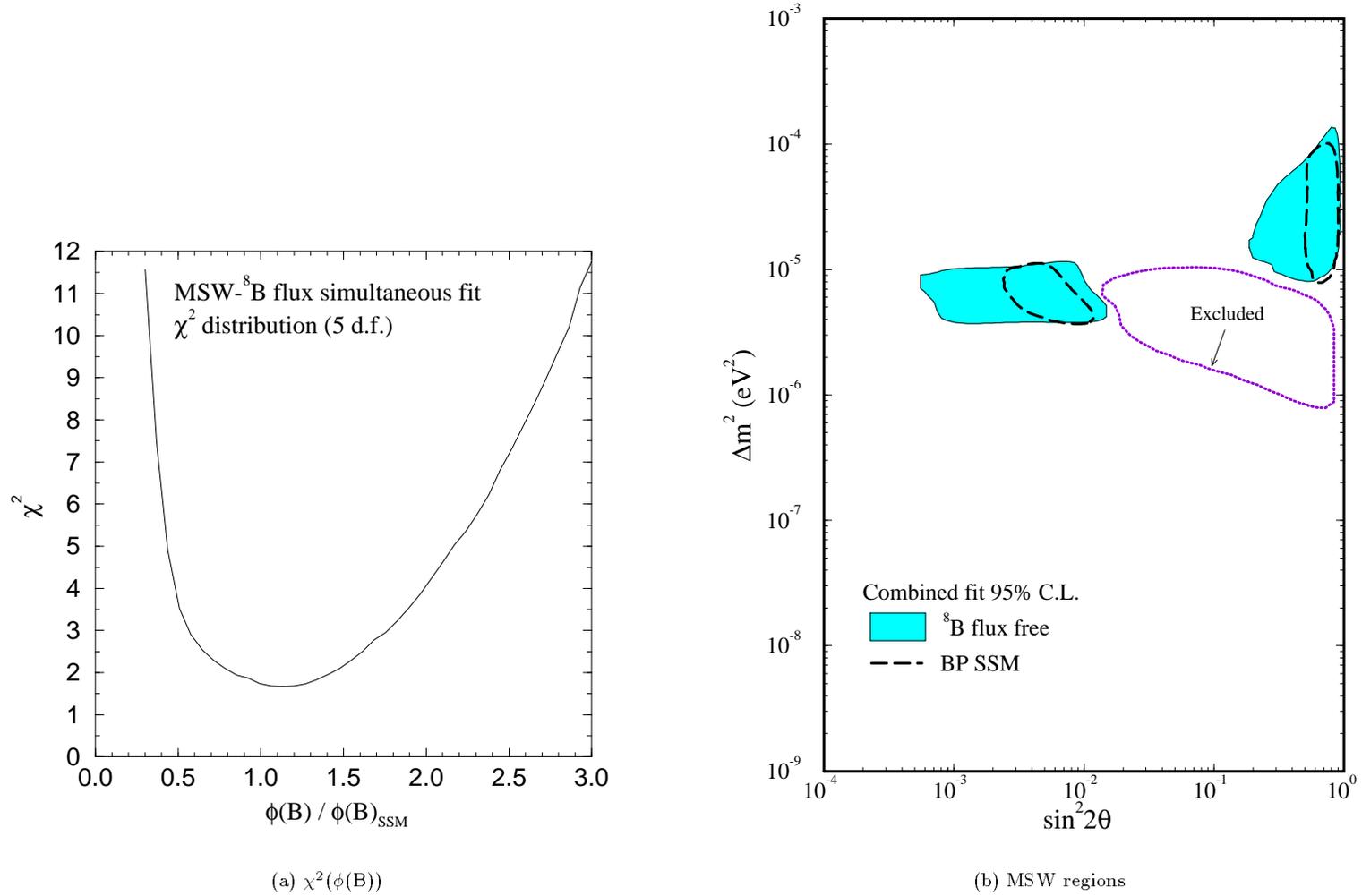


\caption{%
The MSW-$\phi({\rm B})$ simultaneous fit to the existing data.  This
is a 3 parameter fit for 8 data points, including 6 Kamiokande
day-night data bins (5 d.f.)  (a) The $\chi^2$ distribution as a
function of $\phi({\rm B})$.  The current data constrain $\phi({\rm
B})/\phi({\rm B})_{\rm SSM} = 1.15 \pm 0.53$ ($1\sigma$). (b) The MSW
allowed regions.  The corresponding constraints on the $^8$B flux are
displayed in Fig.~\protect\ref{fig:beb_curr_msw}.  There is a third
allowed region around $\sin^22\theta \sim 1$ and $\Delta m^2 \sim 0.5
\times 10^{-7}\; \mbox{eV}^2$, which is too small to be shown in the
figure.  Also shown is the region excluded by the Kamiokande day-night
data (95\% C.L., dotted line), which is independent of the $^8$B flux
uncertainty. For comparison, the allowed regions obtained assuming the
Bahcall-Pinsonneault SSM and its uncertainties are also shown.  }
\label{fig:Bfree_msw}
\end{figure}

Once the MSW effect is introduced in the analysis, the calibration of
the neutrino fluxes becomes more complicated.  One must constrain the
initial fluxes and the MSW parameters simultaneously, while the
neutrino spectrum can be distorted depending on the MSW parameters.
We consider the simplest scenario, the two-flavor MSW effect.
With the three existing data points and using $\Delta m^2$,
$\sin^22\theta$, and $\phi({\rm B})$ as completely free parameters, one
obtains
\begin{equation}
	\phi({\rm B}) / \phi({\rm B})_{\rm SSM}
        =     1.15 \pm 0.53 \;  (1\sigma),
\end{equation}
while the other fluxes are fixed to the SSM values.  Although the
constraint is weak, it is consistent with the SSM predictions and
already excludes (in the MSW context) some of the nonstandard models
with a smaller $^8$B flux.  Since half of the SSM $^8$B flux is seen
in Kamiokande and since the MSW effect only reduces observed rates,
the $^8$B flux cannot be too small.  Taking into account the Homestake
and gallium data and also the neutral current contribution in
Kamiokande, the 90\% lower and upper limit is 0.47 and 2.07 of SSM,
respectively.  The constraint at 90\% C.L.\ is displayed in
Fig.~\ref{fig:beb_curr_msw}.  The $\chi^2$ distribution and the
corresponding constraints on the MSW parameters are shown in
Fig.~\ref{fig:Bfree_msw}.

If the $^7$Be flux is introduced as an additional free parameter, no
realistic constraint is obtained, even though the chlorine and gallium
experiments have sensitivity to the flux.  This is because the MSW
survival probability for the $^7$Be flux can be zero, allowing
essentially any amplitude for the initial flux.  In principle, the
$^7$Be flux has an upper limit due to the luminosity constraint, but
the constraint is weak and irrelevant.  We have repeated the fit by
assuming smaller uncertainties for all experiments and by
incorporating the Kamiokande spectral and day-night data
\cite{Kamiokande-II}, but no constraint was obtained for the $^7$Be
flux.  If MSW is operative, one needs a neutral current sensitivity
for the flux (as in BOREXINO, HELLAZ, and HERON) to extract the $^7$Be
amplitude, which we will discuss in the next section.

\begin{figure}[h]
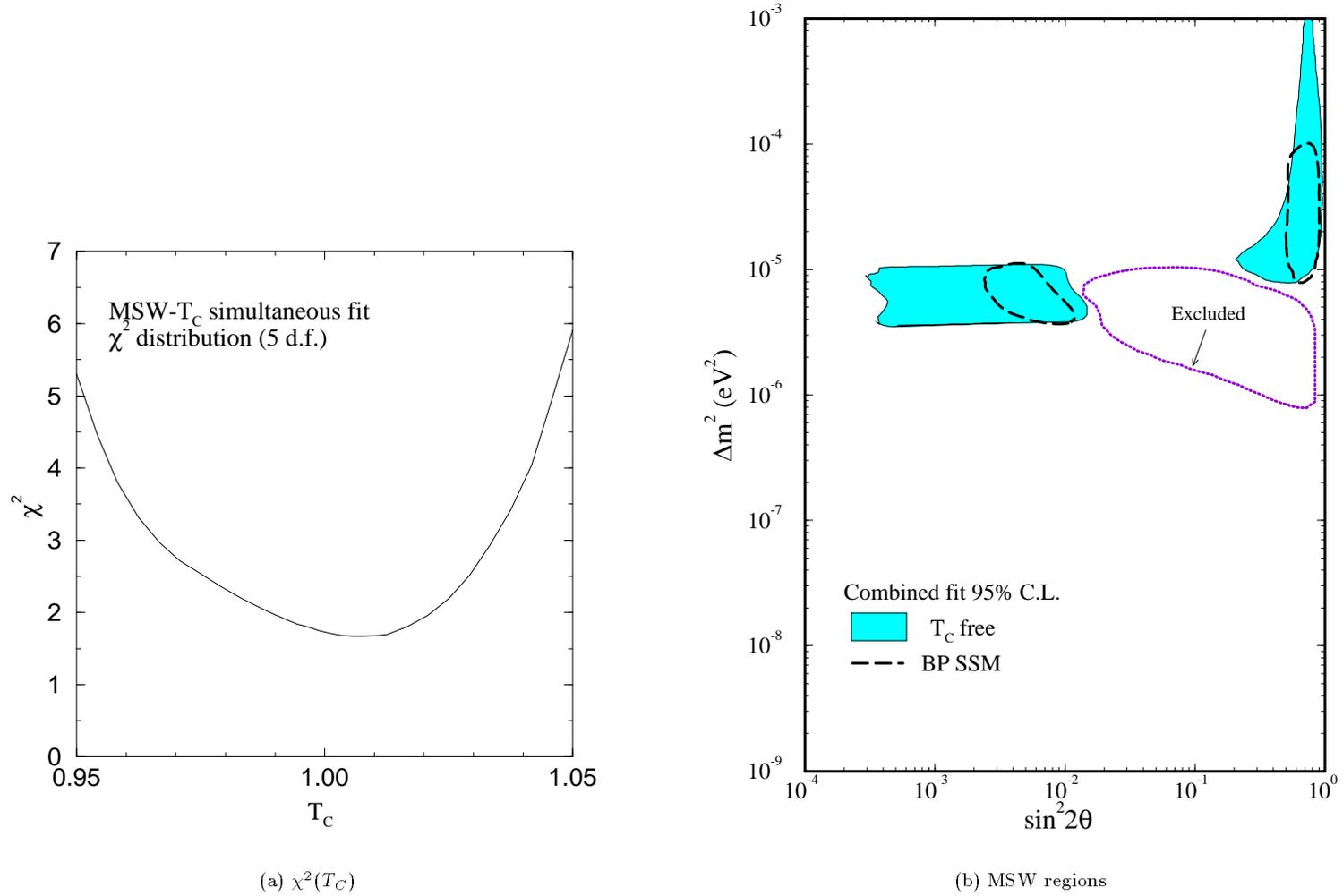


\caption{%
The MSW-$T_C$ simultaneous fit to the existing data.  This is a 3
parameter fit for 8 data points, including 6 Kamiokande day-night data
bins (5 d.f.)  (a) The $\chi^2$ distribution as a function of $T_C$.
The current data constrain $T_C = 1.00 \pm 0.03$, consistent with the
SSM ($T_C = 1 \pm 0.006$).  (b) The MSW allowed regions.  There is a
third allowed region around $\sin^22\theta \sim 1$ and $\Delta m^2
\sim 0.7 \times 10^{-7}\; \mbox{eV}^2$, which is too small to be shown
in the figure.  Also shown is the region excluded by the Kamiokande
day-night data (95\% C.L., dotted line), which is independent of
$T_C$.  For comparison, the allowed regions obtained assuming the
Bahcall-Pinsonneault SSM and its uncertainties are also shown.  }
\label{fig:tcfree_msw}
\end{figure}

The core temperature, although model dependent, can be determined from
the existing data using the power law for the $^7$Be and $^8$B fluxes.
The power law obtained from a Monte Carlo estimation is
\begin{equation}
	\phi({\rm Be}) \sim T_C^{8} \; \mbox{ and } \;
        \phi({\rm B})  \sim T_C^{18},
\end{equation}
and the $T_C$ dependence of the $pp$ flux is obtained from the above
relation and the luminosity constraint (Eq.~\ref{eq:luminosity}),
assuming the exponents of the $pep$ and CNO fluxes as 2.8 and 22,
respectively \cite{HL-MSW-analysis}.  The flux uncertainties from the
nuclear reaction cross sections are included for $S_{17}$ and $S_{34}$
as described in \cite{HL-MSW-analysis}.  The detector cross section
uncertainties for chlorine and gallium are also included.  As a result
of a three parameter fit (two MSW parameters and $T_C$)
\cite{BHKL,HL-MSW-analysis}, we obtain
\begin{equation}
	T_C = 1.00 \pm 0.03 \; (1 \sigma)
\end{equation}
in units of the SSM prediction ($T_C = 1 = 15.57 \times 10^6 \;
\mbox{K}$).  The result is consistent with the SSM ($T_C = 1 \pm 0.006$).
That is, allowing the MSW effect, the present data determine $T_C$ to
within 3\% and are consistent with the SSM predictions.  We note that
without the MSW effect no temperature could describe the data
simultaneously \cite{BKL,BHKL,Shi-Schramm-Dearborn}.  The $\chi^2$
distribution and the constraints on the MSW parameters are shown in
Fig.~\ref{fig:tcfree_msw}.

\subsection{ 		Future                                   }

In the next decade or so, the new generation of solar neutrino
experiments will start and provide high-statistics data.  Those
experiments will measure the fluxes precisely and will allow a
separation of the $^8$B, $^7$Be, and $pp$ fluxes.  Then, to determine
the initial neutrino spectrum in the presence of nonstandard
particle physics effects, what needs to be measured, and with what
accuracy?

We answer these questions quantitatively in the model independent
framework, assuming two-flavor MSW oscillations, since it is the
simplest solution of the solar neutrino problem and most successful in
describing the existing data.
\footnote{
Similar analyses should be applicable to other particle physics
effects if they do not involve too many new parameters.  } %
We assume that the measurement of the charged to neutral current ratio
in SNO will establish neutrino oscillations.  We also assume that the
measurement of the energy spectrum distortions and the day-night
effect in SNO and Super-Kamiokande will distinguish the three separate
MSW parameter branches from each other \cite{HL-MSW-analysis} and from
vacuum oscillations \cite{NH-vacuum}.  The adiabatic and nonadiabatic
regions will show $^8$B spectrum depletion at higher and lower
energies, respectively, which will be observable in SNO and
Super-Kamiokande \cite{HL-MSW-analysis}.  Most of the large-angle
region shows the Earth effect, which will be measurable as day-night
asymmetries or diurnal signal variations in SNO, Super-Kamiokande, and
BOREXINO \cite{HL-MSW-analysis,Baltz-Weneser-94}.  Since the spectrum
and time-variation information constrain the MSW parameters
independent of the flux uncertainties, it would be best to incorporate
those data directly in the analysis once the actual data are
available.  At present, however, we do not attempt to consider such
constraints.  We only consider the averaged SNO NC and BOREXINO rate
(and the averaged Super-Kamiokande rate for some cases) as the minimal
hypothetical data from the future experiments.  Even so, one should be
able to determine all of the parameters reasonably well.

In the MSW calculations, we employ the electron density profile
function and the neutrino production profile functions of the
Bahcall-Pinsonneault SSM.  These functions are solar model dependent
and should, in principle, be an additional source of uncertainties in
constraining the fluxes and the MSW parameters.  We have previously
investigated those uncertainties by using three different SSMs and
also by changing the peak location of the production profiles and the
electron density scale height by 10\% each \cite{HL-MSW-analysis}.
The effect on the obtained MSW parameters was negligible in the
combined fit.

\begin{figure}[h]
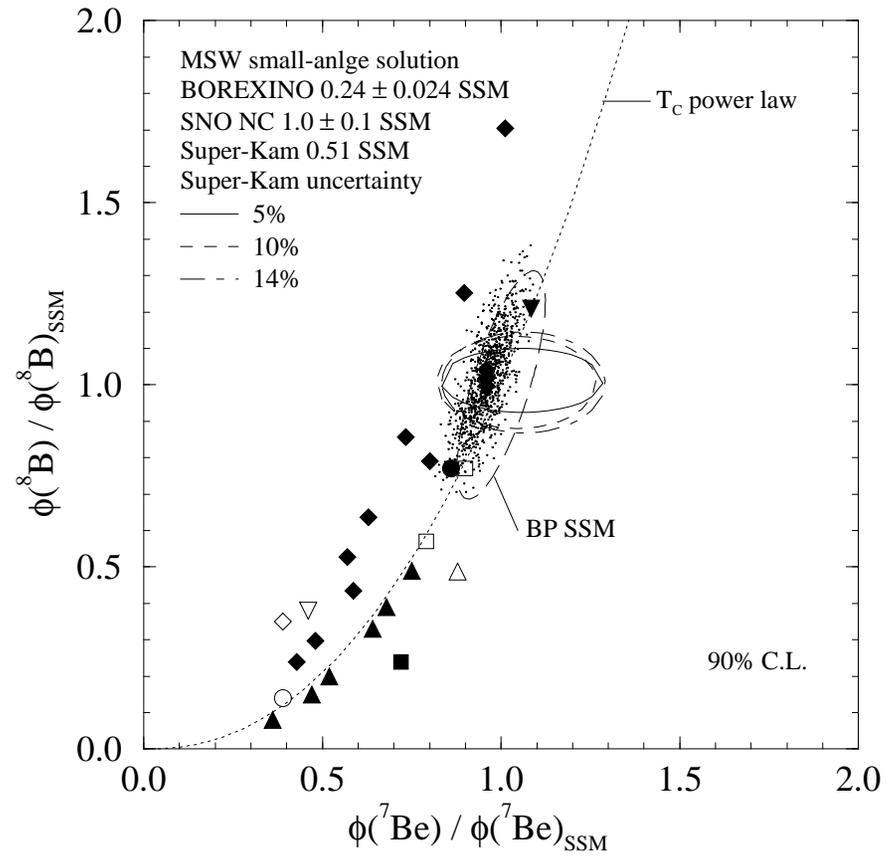


\caption{
The flux constraint when hypothetical Super-Kamiokande results for
various measurement uncertainties are included. The joint fit also
includes the existing data and the hypothetical SNO NC and BOREXINO
results.  The MSW parameters are in the nonadiabatic region.  }
\label{fig:superk_errors}
\end{figure}

In Figures~\ref{fig:each}--\ref{fig:superk_errors}, we consider the
flux constraints for various possible outcomes of the SNO NC,
BOREXINO, and Super-Kamiokande experiments that are consistent with
the assumption that the MSW parameters are in the nonadiabatic
(diagonal) branch.  We include the current results of the Homestake,
(time-averaged) Kamiokande, and the combined gallium experiments,
incorporating the detector cross section uncertainties in the
radio-chemical experiments.  However, omitting either the Homestake or
gallium results does not change the result significantly, which will
allow us to check consistency among data in the future.  The
constraints are obtained by fits to five free parameters [$\phi(pp)$,
$\phi({\rm Be})$, $\phi({\rm B})$, $\Delta m^2$, and $\sin^22\theta$]
imposing the luminosity constraint.  The CNO and the minor fluxes are
fixed to the SSM values.  As shown later, using the CNO flux as an
additional free parameter does not change our results significantly.

Fig.~\ref{fig:each}(a) displays the constraints on the $^7$Be and
$^8$B fluxes at 90\% C.L. when the result of the SNO NC measurement is
assumed to be the SSM value.  The current data from Homestake,
Kamiokande, SAGE, and GALLEX are also included.  The measurement
uncertainties are taken as 10\% of the signal.  The SNO NC rate is
unaffected by flavor oscillations, and yields a direct measurement of
the $^8$B flux.
\footnote{
The charged current (CC) measurement in SNO, combined with the NC
result, will determine the survival probability of the $^8$B flux.
Once neutrino oscillations are established, however, this information
will not significantly improve the Kamiokande result included here.
The effect of the CC measurement uncertainties are similar to the
effect of the Super-Kamiokande measurement uncertainties discussed
below.  }
The $^7$Be flux is not constrained even though the chlorine and
gallium detectors have sensitivity, because the $\nu_e$ survival
probability for this energy range can be zero, and therefore the
initial $^7$Be flux can take essentially any value.

When the $^7$Be measurement from BOREXINO
\footnote{
Our results apply for other $^7$Be measurements with electron
scattering, such as in HELLAZ and HERON. }
is assumed to be $0.24 \pm 0.024$ of the SSM value, the allowed region
is shown in Fig.~\ref{fig:each}(b); the existing data are also
included, but not the SNO result.  Interestingly, both the $^7$Be and
$^8$B fluxes are constrained in this case.  The crucial factor is
that, in electron scattering experiments, there is a contribution of
the neutral current events from the $\nu_\mu$ or $\nu_\tau$ into which
$\nu_e$ oscillates.  For the $^7$Be line spectrum at 0.862 MeV, the
$\nu_\mu - e$ (or $\nu_\tau - e$) cross section is 21\% of the $\nu_e
- e$.  Therefore there should be a signal of at least 21\% of the
initial flux even if the $\nu_e$ survival probability is zero.  The
BOREXINO result, combined with the existing data, gives a stringent
constraint on both the original $^7$Be flux and the MSW parameters,
and those, in turn, constrain the $^8$B flux when combined with the
Kamiokande result.

The results when both the SNO NC and BOREXINO data are assumed for
various different central values are shown in Fig.~\ref{fig:diff}.
The MSW regions for the same or similar SNO NC and BOREXINO results
are displayed in Fig.~\ref{fig:p_fut_diff}.  Considering that the
constraints are independent of solar models, the allowed regions are
determined surprisingly well.  We also note that the information of
the spectral distortion and of the day-night asymmetry, which is
ignored here, will distinguish the adiabatic, nonadiabatic, and
large-angle regions, and therefore further constrain the MSW
parameters.

The effect on the flux constraints for various measurement
uncertainties for SNO NC and BOREXINO are shown in
Fig.~\ref{fig:errors}.  Shown in Fig.~\ref{fig:superk_errors} is the
constraint when the result from Super-Kamiokande is included for
different measurement uncertainties.

\begin{figure}[h]
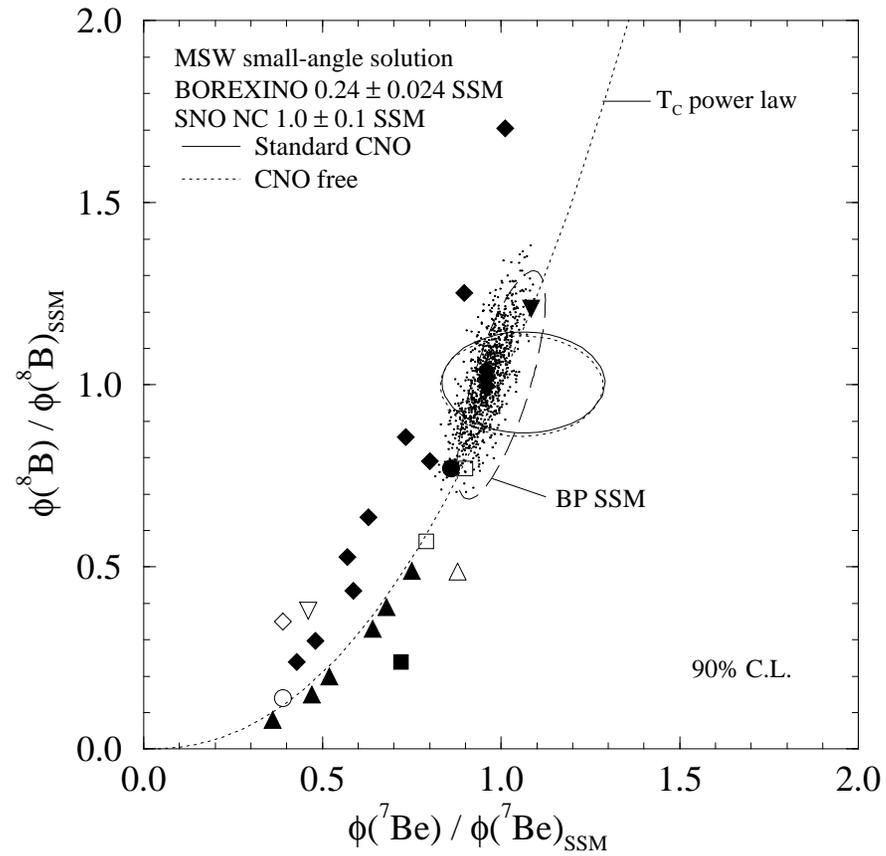


\caption{
The flux constraints when the CNO flux is used as an additional free
parameter.  The joint fit includes the existing data and the
hypothetical SNO NC and BOREXINO results.  The MSW parameters are in
the nonadiabatic region.   }
\label{fig:cnofree}
\end{figure}

The effect of using the CNO flux as an additional free parameter is
displayed in Fig.~\ref{fig:cnofree}; the result is essentially
unchanged.

The analysis has been repeated assuming measurements consistent with
the MSW large-angle branch.  The constraints from each SNO NC and
BOREXINO measurement and the combined SNO NC and BOREXINO are
displayed in Fig.~\ref{fig:la_each} and Fig.~\ref{fig:la_diff}.  The
other results are essentially the same as for the nonadiabatic branch.

{}From the analysis above, we conclude that, if the SNO NC and BOREXINO
uncertainties are at the 10\% level relative to signal, the $^7$Be and
$^8$B fluxes should be constrained at the $\pm$ 20\% and $\pm$ 15\%
level, respectively.  This will clearly distinguish the standard and
nonstandard solar models and perhaps even constrain the SSM
parameters.  The neutral current reaction for the $^7$Be measurement
ensures a non-zero signal (assuming flavor oscillations), which is
especially important for obtaining stringent constraints on the
neutrino fluxes and for distinguishing between competing solar models.

\begin{figure}[h]
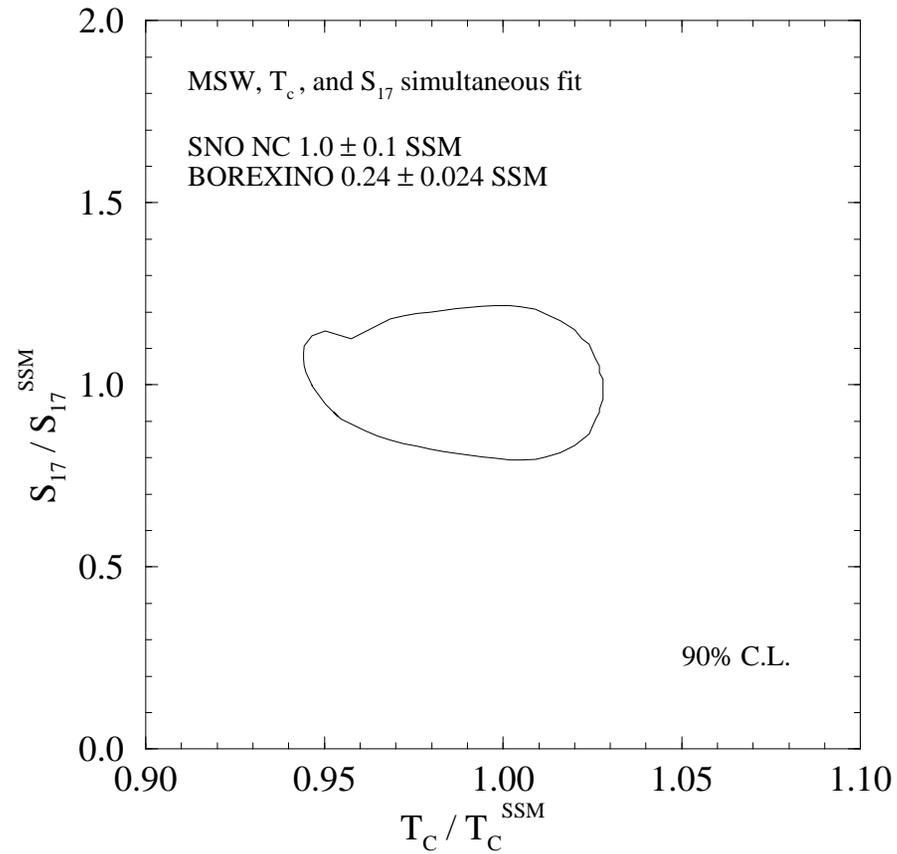


\caption{
The constraints for $T_C$ and $S_{17}$ in the presence of MSW
oscillations when the existing data plus both hypothetical SNO NC and
BOREXINO results are considered.  This is a 4 parameter fit (2 MSW
parameters plus $T_C$ and $S_{17}$) to 5 data points (3 existing data
plus 2 hypothetical results) with the luminosity constraint. }
\label{fig:future_Tc-S17}
\end{figure}

Although model dependent, we have also carried out a simultaneous fit
of $T_C$ and $S_{17}$ with hypothetical outcome from SNO NC and
BOREXINO.  The current data are also included.  The constraint on
$T_C$ and $S_{17}$ is shown in Fig.~\ref{fig:future_Tc-S17}.  $T_C$
and $S_{17}$ will be simultaneously determined at the $\pm 4$\% and
$\pm 20$\% level (90\% C.L.)

\section{		Conclusion	}
\label{sec:conclusion}

We have demonstrated that a model independent analysis using the four
relevant fluxes ($pp$, $^7$Be, $^8$B, and CNO) as free parameters
subject to the luminosity constraint is a feasible scheme for neutrino
spectroscopy, and therefore for testing solar models.  The analysis is
viable with both standard and nonstandard neutrinos.

Assuming standard neutrinos, the existing experiments give a poor fit
and essentially exclude any solar models.  Even allowing this poor
fit, there is no reasonable explanation for the following constraints
from the data: the $^7$Be and CNO fluxes are zero, and the $^8$B flux
is about 40\% of the SSM prediction; the $pp$ flux is the maximum
value allowed by the luminosity constraint:
\begin{eqnarray}
	\phi(pp) / \phi(pp)_{\rm SSM}              & = &  1.089 - 1.095
							         \\
	\phi({\rm Be}) / \phi({\rm Be})_{\rm SSM}  & \le &  0.07
							         \\
	\phi({\rm B})  / \phi({\rm B})_{\rm SSM}   & = &  0.41 \pm 0.04
							         \\
	\phi({\rm CNO})  / \phi({\rm CNO})_{\rm SSM} & \le &  0.26,
\end{eqnarray}
where the uncertainties are at 1$\sigma$; the CNO flux includes the
$^{13}$N and $^{15}$O neutrinos, which are varied with the same scale
factor.  When the constraints are expressed as absolute fluxes, one
obtains
\begin{eqnarray}
	&\phi(pp)	 =   & (6.53 - 6.57) \times 10^{10}
					\; \mbox{cm}^{-2}\mbox{sec}^{-1} \\
	&\phi({\rm Be})	 \le &  0.34 \times 10^{9}
					\; \mbox{cm}^{-2}\mbox{sec}^{-1} \\
	&\phi({\rm B})	 =   & (2.33 \pm 0.23) \times 10^{6}
					\; \mbox{cm}^{-2}\mbox{sec}^{-1} \\
	&\phi({\rm N})	 \le &  1.28 \times 10^{8}
					\; \mbox{cm}^{-2}\mbox{sec}^{-1} \\
	&\phi({\rm O})	 \le &  1.11 \times 10^{8}
					\; \mbox{cm}^{-2}\mbox{sec}^{-1}.
\end{eqnarray}
This severe suppression of the $^7$Be flux relative to the $^8$B flux
is inconsistent with any of the explicit nonstandard solar models.
This problem is made even worse if $S_{17}$ is lower than the values
usually assumed.  Even discarding any one of the three data, the
constraints are consistent with the above.

When the two-flavor MSW effect is introduced in the analysis, the flux
constraint from the current data is weak, but consistent with the SSM,
sufficient to exclude the nonstandard models with too-small $^8$B
fluxes:
\begin{equation}
	\phi({\rm B}) / \phi({\rm B})_{\rm SSM}
        =     1.15 \pm 0.53 \; (1\sigma)
\end{equation}
or
\begin{equation}
	\phi({\rm B})  = (6.54 \pm 3.02) \times 10^{6}
			\mbox{cm}^{-2}\mbox{sec}^{-1} \; (1\sigma)
\end{equation}
No meaningful constraint is obtained if the other fluxes are
introduced as free parameters.

We have also considered the flux constraints in the presence of
two-flavor MSW by assuming various outcomes from the next generation
high-counting experiments.  Of course, one can always consider more
complicated particle physics effects, such as three-flavor
oscillations involving sterile neutrinos.  Here, however, we consider
the simplest scenario, expecting that, should two-flavor MSW be the
case, it will be established as the most likely solution by the NC
measurement in SNO, and by spectral distortions and day-night
asymmetry measurements in SNO and Super-Kamiokande.  Assuming
hypothetical outcomes from the SNO NC and BOREXINO measurements with
realistic uncertainties for this simplest scenario, we found that the
$^7$Be and $^8$B fluxes will be determined at the $\pm$20\% and
$\pm$15\% levels, making competing solar models distinguishable even
if the MSW effect is operative.  The MSW parameters will also be
determined with sufficient accuracy independent of solar models.  We
emphasize that the neutral current sensitivity for $^7$Be neutrinos in
BOREXINO, HELLAZ, and HERON is essential for obtaining such
constraints.  We did not incorporate the information from the SNO CC
rate, Super-Kamiokande rate, spectral distortions, or day-night
asymmetry; those data should provide more stringent constraints on the
MSW parameters as well as on the fluxes.  The $pp$ flux can be
measured by the HELLAZ and HERON experiments, but a measurement
uncertainty at the few \% level is required to determine the flux more
accurately than the luminosity constraint.

\acknowledgements

We thank Eugene Beier and Sidney Bludman for useful discussions.  This
work is supported by the Department of Energy Contract
DE-AC02-76-ERO-3071.

% ***  The end of text  *******************************************************
%******************************************************************************
%
%                        * * * REFERENCE * * *

%******************************************************************************
% Here comes the tables*** ****************************************************

%                        * * * TABLES * * *

\begin{table}[hbt]
\caption{
%
%                                  TABLE I
%
The standard solar model predictions of Bahcall and Pinsonneault (BP SSM)
\protect\cite{Bahcall-Pinsonneault} and of Turck-Chi\'eze and Lopes
(TL SSM) \protect\cite{Turck-Chieze-Lopes}, along with the results of
the solar neutrino experiments.
}
\label{tab:expdata}
\vspace{1.0ex}

\begin{tabular}{l  c c c}
%
%\hline%-----------------------------------------------------------------------
%\hline%-----------------------------------------------------------------------
               & BP SSM        & TL SSM       & Experiments          \\[1ex]
\hline\\[-2ex]%----------------------------------------------------------------
Kamiokande \tablenotemark[1] ($10^6\; \mbox{cm}^{-2}\mbox{sec}^{-1}$)
	       & $5.69\pm 0.82$ & $4.4\pm 1.1$&
				$2.89 \pm 0.41$ (0.51 $\pm$ 0.07 BP SSM) \\
Homestake \tablenotemark[2] (SNU)
	       &  8 $\pm$ 1  & 6.4 $\pm$ 1.4   & 2.55 $\pm$ 0.25
                                                 (0.32 $\pm$ 0.03 BP SSM)\\
SAGE \tablenotemark[3] \& GALLEX \tablenotemark[4] (SNU)
		& 131.5 $^{+7}_{-6}$ & 122.5 $\pm$ 7 & 77 $\pm$ 9
                                                    (0.59 $\pm$ 0.07 BP SSM)
%\hline%-----------------------------------------------------------------------
%\hline%-----------------------------------------------------------------------
\end{tabular}
\vspace{1ex}
\tablenotetext[1]{%
The combined result of Kamiokande II and III (total of 1670 days) is
2.89 +0.22/--0.21 (stat) $\pm$ 0.35 (sys) $\times 10^6$
cm$^{-2}$sec$^{-1}$ \cite{Kamiokande-III}.     }
\tablenotetext[2]{%
The result through June, 1992 (Run 18 -- 124) is
2.55 $\pm$ 0.17 (stat) $\pm$ 0.18 (sys) SNU \cite{Homestake-update}. }
\tablenotetext[3]{%
The combined result of SAGE I and II (through January, 1993) is
74 +13/--12 (stat) +5/--7 (sys) SNU \cite{SAGE-update}. }
\tablenotetext[4]{%
The combined result of GALLEX I and II (30 runs, through October, 1993)
is 79 $\pm$ 10 (stat) $\pm$ 6 (sys) SNU \cite{GALLEX}. }
\end{table}

\vspace{7ex}
\begin{table}[hbt]
\caption{
%
%                                  TABLE II
%
To simplify the notation, we use the following neutrino fluxes as
units.  These reference fluxes correspond to the Bahcall-Pinsonneault
standard solar model with the helium diffusion effect
\protect\cite{Bahcall-Pinsonneault}.
}
\label{tab:BP-fluxes}
\vspace{1.0ex}
\begin{tabular}{l  c c c}
%
%\hline%-----------------------------------------------------------------------
%\hline%-----------------------------------------------------------------------
				& cm$^{-2}$ sec$^{-1}$   \\
\hline\\[-2.5ex]%--------------------------------------------------------------
$\phi(pp)_{\rm SSM} $		& $  6.00 \times 10^{10} $  \\
$\phi(pep)_{\rm SSM}$		& $  1.43 \times 10^{8}  $  \\
$\phi(hep)_{\rm SSM}$		& $  1.23 \times 10^{3}  $  \\
$\phi({\rm Be})_{\rm SSM}$   	& $  4.89 \times 10^{9}  $  \\
$\phi({\rm B})_{\rm SSM}$    	& $  5.69 \times 10^{6}  $  \\
$\phi({\rm N})_{\rm SSM}$ 	& $  4.92 \times 10^{8}  $  \\
$\phi({\rm O})_{\rm SSM}$ 	& $  4.26 \times 10^{8}  $  \\
$\phi({\rm F})_{\rm SSM}$ 	& $  5.39 \times 10^{6}  $  \\
%
%\hline%-----------------------------------------------------------------------
%\hline%-----------------------------------------------------------------------
%
\end{tabular}
\end{table}

%\clearpage
\begin{table}[hbt]
\caption{
%
%                                  TABLE III
%
The constraints on fluxes from various combinations of the current
data with and without the MSW effect.  The uncertainties are at
1$\sigma$, and the fluxes are in units of the reference values defined
in Table~\protect\ref{tab:BP-fluxes}.  The constraints are converted
to absolute fluxes in Table~\protect\ref{tab:flux-constraints-abs}.
The upper limit on the $pp$ flux (1.095) is due to the luminosity
constraint.  Without the MSW effect, we note that the constraints are
consistent with each other even if any one of the three data is
ignored, but are inconsistent with the SSM and nonstandard solar
models, which generally suppress the $^8$B flux more than the $^7$Be
flux.  When the MSW effect is present, a reasonable constraint is
obtained only for the $^8$B flux.  The obtained flux is consistent
with the SSM prediction, albeit with a large uncertainty.
}
\label{tab:flux-constraints}
\vspace{1.0ex}
\begin{tabular}{l  c c c c}
%
%\hline%-----------------------------------------------------------------------
%\hline%-----------------------------------------------------------------------
		& $pp$ 		& $^7$Be	&$^8$B 		& CNO       \\
\hline%------------------------------------------------------------------------
\multicolumn{5}{l}{ Constraints without the MSW effect } \\
Kam + Cl + Ga 	&1.089 -- 1.095 & $< 0.07$	& $0.41\pm0.04$ & $< 0.26$  \\
Kam + Cl 	&1.084 -- 1.095 & $< 0.13$	& $0.42\pm0.04$ & $< 0.38$  \\
Kam + Ga 	&1.085 -- 1.095 & $< 0.13$	& $0.50\pm0.07$ & $< 0.56$  \\
Cl  + Ga 	&1.082 -- 1.095 & $< 0.16$	& $0.38\pm0.05$ & $< 0.72$  \\
\hline%------------------------------------------------------------------------
\multicolumn{5}{l}{ Constraints with the MSW effect} \\
Kam + Cl + Ga 	& $<1.095$	&  ---		& $1.15\pm0.53$ &  ---  \\
%\hline%-----------------------------------------------------------------------
%\hline%-----------------------------------------------------------------------
%
\end{tabular}
\end{table}

\begin{table}[hbt]
\caption{
%
%                                  TABLE IV
%
The same as Table~\protect\ref{tab:flux-constraints}, but in units of
absolute fluxes.  The $^{13}$N and $^{15}$O fluxes are varied with the
same scale factor in the fits.
}
\label{tab:flux-constraints-abs}
\vspace{1.0ex}
\begin{tabular}{l  c c c c}
%
%\hline%-----------------------------------------------------------------------
%\hline%-----------------------------------------------------------------------
		& $pp$ \tablenotemark[1]
				& $^7$Be \tablenotemark[2]
				&$^8$B \tablenotemark[3]
				&$^{13}$N and $^{15}$O \tablenotemark[4]     \\
\hline%------------------------------------------------------------------------
\multicolumn{5}{l}{ Constraints without the MSW effect } \\
Kam + Cl + Ga 	& 6.53 -- 6.57	& $< 0.34$ 	& $2.33 \pm 0.23$
							& $< 1.28$, $< 1.11$ \\
Kam + Cl 	& 6.50 -- 6.57  & $< 0.64$	& $2.39 \pm 0.23$
							& $< 1.87$, $< 1.62$ \\
Kam + Ga 	& 6.51 -- 6.57  & $< 0.64$	& $2.85 \pm 0.40$
							& $< 2.76$, $< 2.39$ \\
Cl  + Ga 	& 6.49 -- 6.57  & $< 0.78$	& $2.16 \pm 0.28$
							& $< 3.54$, $< 3.07$ \\
\hline%------------------------------------------------------------------------
\multicolumn{5}{l}{ Constraints with the MSW effect} \\
Kam + Cl + Ga 	& $< 6.57$	& --- 		& $6.54 \pm 3.02$ & ---
%\hline%-----------------------------------------------------------------------
%\hline%-----------------------------------------------------------------------
%
\end{tabular}
\tablenotetext[1]{In units of $10^{10} \; \mbox{cm}^{-2}\mbox{sec}^{-1}$.  }
\tablenotetext[2]{In units of $10^{9}  \; \mbox{cm}^{-2}\mbox{sec}^{-1}$.  }
\tablenotetext[3]{In units of $10^{6}  \; \mbox{cm}^{-2}\mbox{sec}^{-1}$.  }
\tablenotetext[4]{In units of $10^{8}  \; \mbox{cm}^{-2}\mbox{sec}^{-1}$.  }
\end{table}

%****  Figures ****************************************************************

%                    *            *           *
%
%				Figure
%

%                    *            *           *
%
%				Figure
%

%                    *            *           *
%
%				Figure
%

%                    *            *           *
%
%				Figure
%

%                    *            *           *
%
%				Figure
%

%
%				Figure
%

%
%				Figure
%

%                    *            *           *
%
%				Figure
%

%                    *            *           *
%
%				Figure
%
\begin{figure}[h]

\caption{%
The flux constraints when the hypothetical results from (a) the $^8$B
flux measurement in SNO and Super-Kamiokande and (b) the $^7$Be flux
measurement in BOREXINO (and in HELLAZ and HERON) are considered.
The standard neutrino properties are assumed. }
\label{fig:beb_no-msw_each}
\end{figure}

%                    *            *           *
%
%				Figure
%
\begin{figure}[h]

\caption{
The flux constraints for the combined SNO/Super-Kamiokande and
BOREXINO results.  The standard neutrino properties are assumed. The
constraints are for (a) different SNO/Super-Kamiokande rates and (b)
different BOREXINO rates.  }
\label{fig:beb_no-msw_diff}
\end{figure}

%                    *            *           *
%
%				Figure
%
\begin{figure}[h]

\caption{
The flux constraints for various measurement uncertainties in (a)
SNO/Super-Kamiokande and (b) BOREXINO.  The standard neutrino
properties are assumed.  With the measurement uncertainties at the
10\% level, one can distinguish between standard and nonstandard solar
models and perhaps even constrain the SSM parameters.  }
\label{fig:beb_no-msw_errors}
\end{figure}

%
%                               Figure
%

%                    *            *           *
%
%				Figure
%

%                    *            *           *
%
%				Figure
%

%                    *            *           *
%
%				Figure
%
\begin{figure}[h]

\caption{%
The flux constraints for the MSW nonadiabatic region when the existing
data plus possible results from (a) SNO and (b) BOREXINO are
considered.  The $pp$, $^7$Be, and $^8$B fluxes are fit as free
parameters subject to the luminosity constraint.  These are 5
parameter fits (2 MSW parameters and 3 fluxes) to 4 data points (3
existing data plus 1 future data) with the luminosity constraint. }
\label{fig:each}
\end{figure}

%                    *            *           *
%
%				Figure
%
\begin{figure}[h]

\caption{
The flux constraints when the combined existing data plus possible
results from both SNO and BOREXINO are considered.  The projected
experimental results are motivated by the MSW small-angle
(nonadiabatic) solution.  The constraints are for (a) different SNO NC
rates and (b) different BOREXINO rates.  These are 5 parameter fits (2
MSW parameters and 3 fluxes) to 5 data points (3 existing data plus 2
future results) with the luminosity constraint.}
\label{fig:diff}
\end{figure}

%                    *            *           *
%
%				Figure
%
\begin{figure}[h]

\caption{
The MSW allowed region when the $pp$, $^7$Be, and $^8$B fluxes are fit
as free parameters with the luminosity constraint.  The existing data
plus the results from SNO NC and BOREXINO are used.  We assume (a)
different SNO NC rates with a fixed BOREXINO rate and (b) different
BOREXINO rates with a fixed SNO NC rate.  The constraints for the
fluxes with similar assumptions are shown in
Fig.~\protect\ref{fig:diff} and \protect\ref{fig:la_diff}.  Using the
CNO flux as an additional free parameter does not change the allowed
regions significantly.  }
\label{fig:p_fut_diff}
\end{figure}

%                    *            *           *
%
%				Figure
%
\begin{figure}[h]

\caption{
The flux constraints for the MSW nonadiabatic region for various
measurement uncertainties in (a) SNO and (b) BOREXINO.  The existing
data are also included in the fits.}
\label{fig:errors}
\end{figure}

%                    *            *           *
%
%				Figure
%

%                    *            *           *
%
%				Figure
%

%                    *            *           *
%
%				Figure
%
\begin{figure}[h]

\caption{
Same as Fig.~\protect\ref{fig:each}, but the MSW parameters are in the
large-angle region. }
\label{fig:la_each}
\end{figure}

%                    *            *           *
%
%				Figure
%
\begin{figure}[h]

\caption{
Same as Fig.~\protect\ref{fig:diff}, but the MSW parameters are in the
large-angle region. }
\label{fig:la_diff}
\end{figure}

%                    *            *           *
%
%				Figure
%


\begin{thebibliography}{99}

\bibitem[*]{Present-address}
Present address:
Department of Physics,
The Ohio State University,
Columbus, Ohio 43210.

\bibitem{Homestake}
R.\ Davis, Jr. {\it et al.},
in {\it Proceedings of the 21th International Cosmic Ray Conference},
edited by R.\ J.\ Protheroe (University of Adelaide Press, Adelaide, 1990),
Vol. 12, p. 143;
%
R.\ Davis, Jr.,
in {\it Frontiers of Neutrino Astrophysics},
edited by  Y.\ Suzuki and K.\ Nakamura (Universal Academy Press, Tokyo, 1993),
P.\ 47.

\bibitem{Homestake-update}
K.\ Lande,
in {\it Neutrino 94}, Eilat, Israel, May-June 1994.

\bibitem{Kamiokande-II}
Kamiokande II Collaboration, K.\ S.\ Hirata {\it et al.},
Phys.\ Rev.\ Lett.\ {\bf 65}, 1297 (1990);
{\bf 65}, 1301(1990); {\bf 66}, 9 (1991);
Phys.\ Rev.\ D {\bf 44}, 2241 (1991).

\bibitem{Kamiokande-III}
Y.\ Suzuki,
in {\it Neutrino 94}, Eilat, Israel, May-June 1994.

\bibitem{SAGE}
SAGE Collaboration, A.\ I.\ Abazov, {\it et al.},
Phys.\ Rev.\ Lett.\ {\bf 67}, 3332 (1991);
%
SAGE Collaboration, J.\ N.\ Abdurashitov {\it et al.},
Phys.\ Lett.\ B {\bf 328}, 234 (1994).

\bibitem{SAGE-update}
V.\ N.\ Gavrin,
in {\it Neutrino 94}, Eilat, Israel, May-June 1994;
SAGE Collaboration, J.\ N.\ Abdurashitov {\it et al.},
to be published in {\it Proceedings of the 5th Conference on the
Intersection of Particle and Nuclear Physics},
St.\ Petersburg, Florida, May-June 1994.

\bibitem{GALLEX}
GALLEX Collaboration, P.\ Anselmann {\it et al.},
Phys.\ Lett.\ B {\bf 285}, 376 (1992);
{\bf 285}, 390 (1992);
{\bf 314}, 445 (1993);
{\bf 327}, 337 (1994).

\bibitem{Bahcall-Pinsonneault}
J.\ N.\ Bahcall and M.\ H.\ Pinsonneault,
Rev.\ Mod.\ Phys.\ {\bf 64}, 885 (1992).

\bibitem{Turck-Chieze-Lopes}
S.\ Turck-Chi\`eze and I.\ Lopes,
Astrophys. J. {\bf 408}, 347 (1993).
S.\ Turck-Chi\`{e}ze, S.\ Cahen, M.\ Cass\'{e}, and C.\ Doom,
Astrophys.\ J.\ {\bf 335}, 415 (1988).

\bibitem{Bahcall-Bethe}
J.\ N.\ Bahcall and H.\ A.\ Bethe,
Phys.\ Rev.\ D {\bf 47}, 1298 (1993);
Phys.\ Rev.\ Lett.\ {\bf 65}, 2233 (1990);
H.\ A.\ Bethe and J.\ N.\ Bahcall,
Phys.\ Rev.\ D {\bf 44}, 2962 (1991).

\bibitem{HBL}
N.\ Hata, S.\ Bludman, and P.\ Langacker,
Phys.\ Rev.\ D {\bf 49}, 3622 (1994).

\bibitem{Bahcall-spectrum}
J.\ N.\ Bahcall,
Phys.\ Rev.\ D {\bf 44}, 1644 (1991).

\bibitem{MSW}
L.\ Wolfenstein,
Phys.\ Rev.\ D {\bf 17}, 2369 (1978); {\bf 20}, 2634 (1979);
%
S.\ P.\ Mikheyev and A.\ Yu.\ Smirnov,
Yad.\ Fiz.\ {\bf 42}, 1441 (1985)
[Sov.\ J.\ Nucl.\ Phys. {\bf 42}, 913 (1985)];
Nuovo Cimento {\bf 9C}, 17 (1986).

\bibitem{HL-MSW-analysis}
N.\ Hata and P.\ Langacker,
Phys.\ Rev.\ D {\bf 50}, 632 (1994).

\bibitem{RIKEN}
T.\ Motobayashi {\it et al.},
Report No.\ Rikkyo RUP 94-2, Yale 40609-1141
(submitted to Phys.\ Rev.\ Lett.)

\bibitem{Johnson-etal}
C.\ W.\ Johnson, E.\ Kolbe, S.\ E.\ Koonin, and K.\ Langanke,
Astrophys. J. {\bf 392}, 320 (1992).

\bibitem{SNO}
G.\ T.\ Ewan {\it et al.}
``Sudbury Neutrino Observatory Proposal'', Report No.\ SNO-87-12, 1987
(unpublished);
``Scientific and Technical Description of the Mark II SNO Detector'',
edited by E.\ W.\ Beier and D.\ Sinclair, Report No.\ SNO-89-15, 1989
(unpublished).

\bibitem{Super-Kamiokande}
Y.\ Totsuka,
University of Tokyo Report No.\ ICRR-Report-227-90-20, 1990
(unpublished).

\bibitem{ICARUS}
J.\ P.\ Revol,
in {\it Frontiers of Neutrino Astrophysics}, edited by Y.\ Suzuki and
K.\ Nakamura (University Academy Press, Inc., Tokyo, Japan, 1993), p.~167.

\bibitem{BOREXINO}
``BOREXINO at Gran Sasso --- proposal for a real time detector for
low energy solar neutrinos'', Vol.\ 1,
edited by G.\ Bellini, M.\ Campanella, D.\ Giugni, and R.\ Raghavan (1991).

\bibitem{HELLAZ}
J.\ Seguinot, T.\ Ypsilantis, and A.\ Zichini,
College de France Report No.\ LPC92-31, 1992 (unpublished).

\bibitem{HERON}
S.\ R.\ Bandler {\it et al.},
Journal of Low Temperature Phys.\ {\bf 93}, 785 (1993).

\bibitem{Bahcall-Ulrich}
J.\ N.\ Bahcall and R.\ N.\ Ulrich,
Rev.\ Mod.\ Phys.\ {\bf 60}, 297 (1988).

\bibitem{Bahcall-book}
J.\ N.\ Bahcall,
{\it Neutrino Astrophysics}, (Cambridge University Press, Cambridge,
England, 1989).

\bibitem{BKL}
S.\ Bludman, D.\ Kennedy, and P.\ Langacker,
Phys.\ Rev.\ D {\bf 45}, 1810 (1992);
Nucl.\ Phys.\ B {\bf 374}, 373 (1992).

\bibitem{BHKL}
S.\ Bludman, N.\ Hata, D.\ Kennedy, and P.\ Langacker,
Phys.\ Rev.\ D {\bf 47}, 2220 (1993).

\bibitem{Castellani-etal-PRD}
V.\ Castellani {\it et al.},
INFN preprint INFNFE-3-94, 1994 (unpublished).

\bibitem{Castellani-Degl'Innocenti-Fiorentini-AA}
V.\ Castellani, S.\ Degl'Innocenti, and G.\ Fiorentini,
Astron.\ Astrophys.\ {\bf 271}, 601 (1993).

\bibitem{Aufderheide-etal}
M.\ B.\  Aufderheide, S.\ D.\ Bloom, D.\ A.\ Resler, and C.\ D.\ Goodman,
Phys.\ Rev.\ C {\bf 49}, 678 (1994).

\bibitem{Bahcall-Ulrich-nonssm-table}
J.\ Bahcall, Ref.~\cite{Bahcall-book}, Table~6.6, p.\ 163.

\bibitem{Dearborn}
D.\ Dearborn,
private communications.

\bibitem{WIMPs}
J.\ Faulkner and R.\ L.\ Gilliland,
Astrophys.\ J.\ {\bf 299}, 994 (1985);
R.\ L.\ Gilliland, J.\ Faulkner, W.\ H.\ Press, and D.\ N.\ Spergel,
Astrophys.\ J.\ {\bf 306}, 703 (1986).

\bibitem{Castellani-Degl'Innocenti-Fiorentini-largeS11}
V.\ Castellani, S.\ Degl'Innocenti, and G.\ Fiorentini,
Phys.\ Lett.\ B {\bf 303}, 68 (1993).

\bibitem{Mixing-model}
R.\ Sienkiewicz, J.\ N.\ Bahcall, and B.\ Paczy\'nski,
Astrophys.\ J.\ {\bf 349}, 641 (1990).

\bibitem{Dar-Shaviv}
A.\ Dar and G.\ Shaviv,
Technion preprint, 1994 (unpublished).

\bibitem{Merryfield}
W.\ Merryfield,
in {\it Solar Modelling}, edited by A.\ B.\ Balantekin and J.\ N.\ Bahcall
(World Scientific, Singapore, 1994).

\bibitem{Gough-Toomre}
D.\ Gough and J.\ Toomre,
Annu. Rev. Astron. Astrophys. {\bf 29}, 627 (1991);
D.\ Gough,
Phil.\ Trans.\ R.\ Soc.\ Lond.\ A {\bf 346}, 37 (1994).

\bibitem{Shi-Schramm-Dearborn}
X.\ Shi, D.\ N.\ Schramm, and D.\ S.\ P.\ Dearborn,
Fermilab Report No.\ FERMILAB-PUB-94-122-A
(Los Alamos Electronic Preprint No.\ astro-ph/9404006).

\bibitem{NH-vacuum}
N.\ Hata,
University of Pennsylvania Report No.\ UPR-0605T, 1994 (unpublished).

\bibitem{Baltz-Weneser-94}
A.\ J.\ Baltz and J.\ Weneser,
Brookhaven National Laboratory Report No.\ BNL-60387, 1994
(to be published in Phys.\ Rev.\ D).

\end{thebibliography}
\end{document}